\documentclass[%
 reprint,
 amsmath,amssymb,
 aps,
nofootinbib]{revtex4-2}

\usepackage{graphicx}
\usepackage{dcolumn}
\usepackage{bm}\usepackage{natbib}
\usepackage{mathrsfs}
\usepackage{latexsym,amsfonts,amsmath,amssymb}
\usepackage{hyperref}
\usepackage{bbm}
\usepackage{empheq}
\usepackage{graphicx}
\usepackage{color}
\usepackage{caption}
\usepackage{subcaption}
\usepackage[normalem]{ulem}
\usepackage{comment}
\usepackage{url}
\usepackage{slashed}
\usepackage{extarrows}
\usepackage{amsmath}
\usepackage{mathtools}
\usepackage{float}
\usepackage{cancel}
\usepackage{amsxtra,graphics,epsfig,bm,tikz,xfrac}
\usepackage{array}
\newcommand{\NN}{\mathcal{N}}

\begin{document}

\preprint{APS/123-QED}
\title{Supersymmetric truncation of {$\mathcal{N}=3$} dilaton Weyl multiplet}
\author{Soumya Adhikari} 
\email{ssoumya.a012@gmail.com}
\affiliation{
Indian Institute of Science Education and Research Thiruvananthapuram,
	Vithura, Thiruvananthapuram, India\\}\affiliation{Department of Physics \& Center for Quantum Spacetime, Sogang University, 35
Baekbeom-ro, Mapo-gu, Seoul 04107, Republic of Korea}

\author{Aravind Aikot} 
\email{arvd1719@alumni.iisertvm.ac.in}
\affiliation{
Indian Institute of Science Education and Research Thiruvananthapuram,\\
	Vithura, Thiruvananthapuram, India\\}
\affiliation{Lehigh University, PA 18015, United States}
\author{Eswar Krishnan} 
\email{eswarkrishnan20@iisertvm.ac.in}
\affiliation{
Indian Institute of Science Education and Research Thiruvananthapuram,\\
	Vithura, Thiruvananthapuram, India}

\author{Bindusar Sahoo} 
\email{bsahoo@iisertvm.ac.in}
\affiliation{
Indian Institute of Science Education and Research Thiruvananthapuram,\\
	Vithura, Thiruvananthapuram, India}

\author{Sharun Shanmughan} 
\email{sharun22@alumni.iisertvm.ac.in}
\affiliation{
Indian Institute of Science Education and Research Thiruvananthapuram,\\
	Vithura, Thiruvananthapuram, India}

\begin{abstract} 
We perform all possible supersymmetric truncations of the {four-dimensional} $\mathcal{N}=3$ dilaton Weyl multiplet, {which realizes} an R-symmetry $SU(2) \times U(1) \times U(1)$, to $\NN=2$. A particular truncation {procedure} does not break any of the R-symmetries and leads to the {known} $\NN=2$ vector-dilaton Weyl multiplet and the $\NN=2$ vector multiplet. A different truncation procedure breaks the $SU(2)$ part of the $R$-symmetry to $U(1)$ and leads to a $32+32$ off-shell representation of  $\mathcal{N}=2$ conformal supergravity {with a partially broken R-symmetry}. Independently, we construct another $32+32$ off-shell multiplet in $\NN=2$ conformal supergravity by coupling the $\mathcal{N}=2$ scalar-tensor multiplet to the $\mathcal{N}=2$ standard Weyl multiplet {and using the scalar fields present in the scalar-tensor multiplet to break the $SU(2)$ R-symmetry to $U(1)$}. We then establish the equivalence between these two multiplets through a mapping. We observe that this $32+32$ multiplet is gauge equivalent to a Poincaré supergravity multiplet as it has all the compensators necessary to go from conformal supergravity to Poincar{\'e} supergravity.
\end{abstract}

\maketitle


\section{\label{sec:level1}Introduction}

Supergravity has been extensively studied for nearly half a century, driven by the strong motivation to bridge the principles of general theory of relativity and quantum field theory. As the supersymmetric theory of gravity, supergravity emerges as the low-energy effective description of superstring theory. The construction of matter-coupled supergravity theories, both with and without higher-derivative terms, has been significantly advanced by the development of the superconformal framework.
In particular, superconformal tensor calculus has played a crucial role in these constructions, wherein {some} Poincaré supergravity theories can be formulated as broken superconformal theories.\footnote{{Conformal supergravity theories are possible in space-time dimensions less than or equal to six and with total number of supersymmetries less than or equal to 32, that includes both ordinary as well as special supersymmetry. Only in these cases one can formulate Poincar{\'e} supergravity theories as broken superconformal theories.}} The conformal approach, with its enlarged symmetry structure, offers several advantages over the direct Poincaré supergravity formulation, providing greater clarity and systematic organization of the theory’s components. This approach was first initiated in the $\mathcal{N}=1$ setting through the pioneering works of S. Ferrara, M. Kaku, P.K. Townsend, and P. van Nieuwenhuizen \cite{Ferrara:1977ij, Kaku:1978nz, Kaku:1978ea}, which was later generalized to $\mathcal{N}=2$ theories in $4$, $5$ and $6$ dimensions \cite{Lauria:2020rhc}.

The superconformal framework offers an off-shell formulation as well, which has been extensively utilized in supersymmetric localization and black hole entropy calculations. In this framework, the field representation of the superconformal algebra is given by the Weyl multiplet, which contains the graviton, gravitino, and all other gauge fields associated with the superconformal symmetries.
In extended supergravity theories, the Weyl multiplet additionally includes certain covariant matter fields necessary for the off-shell closure of the algebra. It is evident that Poincaré supergravity contains more degrees of freedom than conformal supergravity; these are supplied by compensating multiplets, which also serve to gauge-fix the additional symmetries of the conformal theory. For example, in the $\mathcal{N}=3$ theory {in four dimensions}, one requires three compensating vector multiplets to obtain the corresponding Poincaré supergravity from conformal supergravity \cite{Hegde:2022wnb}. On the other hand, in the $\mathcal{N}=2$ theory, there is more flexibility where one can choose among different pairs of compensators. One of the compensators is always the vector multiplet, and it can be paired with the linear multiplet, the nonlinear multiplet, or the hypermultiplet. These different choices lead to distinct formulations of $\mathcal{N}=2$ Poincaré supergravity in four dimensions \cite{deWit:1980lyi}.

Depending on the choice of the covariant fields in the Weyl multiplet, we can obtain different representations of the superconformal algebra. The two most commonly known Weyl multiplets are the standard Weyl multiplet and the dilaton Weyl multiplet. The dilaton Weyl multiplet differs from the standard Weyl multiplet due to the presence of a physical scalar that transforms non-trivially under dilatation transformations.
There are several methods for constructing a dilaton Weyl multiplet. The standard method involves coupling the standard Weyl multiplet with one or more on-shell compensating matter multiplets. This coupling leads to the replacement of some of the auxiliary fields in the standard Weyl multiplet by dynamical fields from the compensators. The replacement occurs by reinterpreting the equations of motion for the compensator fields as constraints on the standard Weyl fields, thereby making some of them composite.
The dilaton Weyl multiplet has been constructed in dimensions less than or equal to six, accommodating various amounts of supersymmetry \cite{Adhikari:2023tzi, Adhikari:2024esl, Nahm:1977tg,Fujita:2001kv, Bergshoeff:2001hc,Bergshoeff:1985mz,Butter:2017pbp, Hutomo:2022hdi, Gold:2022bdk, Ciceri:2024xxf}. Notably, the dilaton Weyl multiplet, in contrast to the standard Weyl multiplet, requires fewer compensators to produce Poincaré supergravity. This is not surprising since the dilaton Weyl multiplet is already obtained by combining one or more compensating multiplets with the standard Weyl multiplet.
In addition, the dilaton Weyl multiplet naturally incorporates a two-form field while remaining off-shell, which facilitates the construction of higher-derivative theories {that may descend} from string theory. Another interesting feature of the dilaton Weyl formulation is that the electromagnetic duality symmetry can be realized, at least partially, in an off-shell manner. In {four-dimensional} $\NN=2$ conformal supergravity there are two different variants of dilaton Weyl multiplet. One is constructed by coupling the vector multiplet to the standard Weyl multiplet \cite{Butter:2017pbp} whereas the other is constructed by coupling the hypermultiplet to the standard Weyl multiplet \cite{Gold:2022bdk}. We will refer to the former as vector-dilaton Weyl multiplet and the latter as the hyper-dilaton Weyl multiplet.

Multiplets with a higher number of supersymmetries can be related to those with fewer supersymmetries through a procedure known as supersymmetric truncation. In this procedure, the multiplets of $\mathcal{N}+1 $ conformal supergravity are expressed in terms of the multiplets of $\mathcal{N}$ conformal supergravity along with the gravitino multiplet. The gravitino multiplet, together with the supersymmetry parameters corresponding to the additional supersymmetries, is set to zero during the truncation. Supersymmetric truncation has been effectively used to relate $\mathcal{N}=2$ and $\mathcal{N}=1$ multiplets \cite{Yamada:2019ttz}, $\mathcal{N}=3$ and $\mathcal{N}=2$ multiplets \cite{Aikot:2024cne} and $\mathcal{N}=4$ and $\mathcal{N}=3$ multiplets \cite{Hegde:2022wnb, vanMuiden:2017qsh}.

An $\mathcal{N}=3$ dilaton Weyl multiplet with $SU(2)\times U(1) \times U(1)$ R-symmetry was recently constructed by coupling an $\mathcal{N}=3$ vector multiplet to the $\mathcal{N}=3$ standard Weyl multiplet in the usual fashion \cite{Adhikari:2024qxg}. In this construction, certain auxiliary fields of the standard Weyl multiplet are replaced with dynamical fields from the vector multiplet.
In this work, we investigate a supersymmetric truncation of this $\mathcal{N}=3$ dilaton Weyl multiplet and explore the structure of the corresponding $\mathcal{N}=2$ multiplet.

Since the R-symmetry present in the $\NN=3$ dilaton Weyl multiplet is $SU(2)\times U(1)\times U(1)$, there are a couple of distinct supersymmetric truncations to the $\NN=2$ theory that one can perform. One of the ways will not break any R-symmetry and will lead to a vector-dilaton Weyl multiplet and a vector multiplet in $\NN=2$ conformal supergravity. The second way will break the $SU(2)$ part of the R-symmetry to $U(1)$ and will lead to an $\NN=2$ multiplet that realizes an $U(1)\times U(1)\times U(1)$ local symmetry and constitutes an off-shell $32+32$ irreducible representation that corresponds to neither a dilaton Weyl nor a standard Weyl.

One can also construct a $32+32$ off-shell multiplet in $\NN=2$ conformal supergravity by coupling the $\mathcal{N}=2$ scalar-tensor multiplet \cite{Aikot:2024cne} with the $\mathcal{N}=2$ standard Weyl multiplet and breaking the $SU(2)$ R-symmetry to $U(1)$ by imposing suitable gauge-fixing conditions on the scalars from the scalar-tensor multiplet. One can also show that the multiplet obtained by the second type of supersymmetric truncation procedure on the $\NN=3$ dilaton Weyl multiplet, mentioned in the previous paragraph, matches with this multiplet. {One can also show that this multiplet is gauge equivalent to a Poincar{\'e} supergravity multiplet as it already has all the necessary compensating fields for gauge fixing the extra symmetries present in conformal supergravity thereby allowing for a completely off-shell formulation of four dimensional $\NN=2$ Poincar{\'e} supergravity. This is a curious aspect of supersymmetric truncation of dilaton Weyl multiplets that constitutes one of the main results of our paper.}

The paper is organized as follows. In section-\ref{dilsusytrunc}, we carry out all possible supersymmetric truncations of the $\NN=3$ dilaton Weyl multiplet with $SU(2)\times U(1)\times U(1)$ R-symmetry \cite{Adhikari:2024qxg} whose component structure is discussed in appendix-\ref{N3dilaton}. In subsection-\ref{susytrunc1}, we use a truncation procedure that does not break any of the R-symmetries and we show that such a truncation procedure leads to the $\NN=2$ vector-dilaton Weyl multiplet and vector multiplet already known in the literature \cite{Butter:2017pbp,deWit:1980lyi} (relevant details are given in Appendix-\ref{N2review}). In subsection-\ref{susytrunc2}, we use a different truncation procedure that breaks the $SU(2)$ part of the R-symmetry to $U(1)$. This leads to a $32+32$ off-shell represntation of $\NN=2$ conformal supergravity that realizes an $U(1)\times U(1)\times U(1)$ local symmetry. In section-\ref{N2const}, we construct an off-shell multiplet in $\NN=2$ conformal supergravity theory by coupling a scalar tensor multiplet with the standard Weyl multiplet and using scalar fields from the scalar tensor multiplet to break the $SU(2)$ part of the R-symmetry to $U(1)$. The resulting multiplet has $32+32$ componens and realizes an $U(1)\times U(1)\times U(1)$ local symmetry. The relevant details of the $\NN=2$ standard Weyl multiplet and the scalar tensor multiplet are given in Appendix-\ref{N2review}. In section-\ref{Relation}, we find explicit mapping that shows the equivalence of the multiplets constructed in subsection-\ref{susytrunc2} and section-\ref{N2const}. In the concluding section, we summarize our results and highlight potential directions for future research.

{The notations and conventions used in the paper are explained in Appendix-\ref{Notation}.}

\section{Supersymmetric truncation from $\NN=3$ to $\NN=2$}\label{dilsusytrunc}
The discovery of $\NN=3$ conformal supergravity in four dimensions is relatively recent compared to its $\NN=2$ and $\NN=4$ counterparts. The development began with the identification of the $\NN=3$ standard Weyl multiplet \cite{vanMuiden:2017qsh,Hegde:2018mxv}, followed by the construction of an invariant action \cite{Hegde:2021rte} resulting in the formulation of higher-derivative $\NN=3$ Poincar{\'e} supergravity \cite{Hegde:2022wnb}. Recently, an $\NN=3$ dilaton Weyl multiplet was discovered \cite{Adhikari:2024qxg} by following the standard procedure, which is as follows. A standard Weyl multiplet was coupled to a vector multiplet, and subsequently, the vector multiplet's field equations were solved algebraically for some of the auxiliary fields of the standard Weyl multiplet. Solving the equations was facilitated by first breaking the $R$-symmetry from $SU(3)$ to $SU(2)\times U(1)$ by using the scalars of the vector multiplet. This led to an $\NN=3$ dilaton Weyl multiplet where the local $R$-symmetry realized is $SU(2)\times U(1)\times U(1)$. The components of the multiplet, together with their transformation rules, can be found in \cite{Adhikari:2024qxg}. However, for the sake of completeness and since we will be using the results, we also give the details in appendix-\ref{N3dilaton}.
For the $\NN=3$ dilaton Weyl multiplet discussed above, the ordinary or $Q$-supersymmetry is parametrized by two sets of supersymmetry parameters. One of them, which is denoted by an $\epsilon^i$, where $i=1,2$, is a doublet of the underlying $SU(2)$ $R$-symmetry whereas the other parameter which we denote as $\epsilon$ is a singlet of the underlying $SU(2)$ $R$-symmetry.\footnote{The left and right chiral components of the $SU(2)$ doublet spinors are denoted using the chiral notation whereas for $SU(2)$ singlet spinors, we explictly use the subscript $L$ and $R$ to denote them. For example the parameter $\epsilon^i$ and $\epsilon_i$ denote the left and right chiral components of the $SU(2)$ doublet supersymmetry parameter whereas $\epsilon_L$ and $\epsilon_R$ refers to the left and right chiral components of the $SU(2)$ singlet parameter.} The same goes for the parameters $\eta^i$ and $\eta$ parameterizing the special or $S$-supersymmetry. There are two distinct ways of performing supersymmetry truncation of this multiplet which we discuss in two separate subsections below. 
\subsection{Supersymmetric truncation-I}\label{susytrunc1}
One obvious way to truncate the $\NN=3$ dilaton Weyl multiplet is to set $\epsilon$ and the corresponding gravitino $\psi_\mu$ to zero as shown below 
\begin{align}\label{trunc_cond_1}
\epsilon_L=0=\psi_{\mu \, L}\;, \; \epsilon_R=0=\psi_{\mu \, R}
\end{align}
This choice will not break any $R$-symmetry and would eventually lead to an $\NN=2$ dilaton Weyl multiplet \cite{Butter:2017pbp} and $\NN=2$ vector multiplet \cite{deWit:1980lyi} known in the literature.

By subsequent applications of supersymmetry transformations on the truncation condition mentioned above \eqref{trunc_cond_1}, we find that the following fundamental field components of the multiplet (together with their conjugates) gets truncated out.
\begin{align}
{T}_{ab}^{i} = \Lambda_{L} =Y_{\mu}^{i}=\chi_{ij}=E_i= \theta_L=\psi_R=0
\end{align}
As a consequence of all the above mentioned truncation, we find that the following composite fields of the $\NN=3$ dilaton Weyl multiplet (See Appendix-\ref{N3dilaton}) will also get truncated out.
\begin{align}
\mathring{\zeta}_{R} =\mathring{D}^i=\mathring{\chi}_{L}=0
\end{align}
\begin{widetext}
Along with that, the S-supersymmetry parameter $\eta$ and the corresponding gauge field $\phi_{\mu}$ will also get truncated out. The supersymmetry transformations of the remaining independent fields  are as given below:
\allowdisplaybreaks
\begin{subequations}\label{2dilatontrans}
\begin{align}
\delta e_\mu^a&=\bar{\epsilon}_i\gamma^a\psi_\mu^i +\text{h.c.} \\[5pt]
\delta\psi_\mu^i&=2\mathcal{D}_\mu\epsilon^i+\frac{1}{8} \varepsilon^{ij} \gamma\cdot \mathring{T}^-\gamma_\mu\epsilon_j-\gamma_\mu \eta^i \\[5pt]
     \delta v_\mu{}^i{}_j&= \bar{\epsilon}^i\phi_{\mu j} -\frac{1}{48} \bar{\epsilon}^i \gamma_\mu \zeta_j +\frac{1}{16}\bigg(\varepsilon_{jk}\bar{\epsilon}^k\gamma_\mu \mathring{\chi}^i\bigg)-\bar{\psi}_\mu^i\eta_j-\text{h.c.}-\text{trace}\\[5pt]
    \delta b_\mu&=\frac{1}{2}\bigg(\bar{\epsilon}^i\phi_{\mu i}-\bar{\psi}_\mu^i\eta_i \bigg)+\text{h.c.} \\[5pt]
    \delta A_\mu&= \frac{i}{6}\bigg( \bar{\epsilon}^i\phi_{\mu i}\bigg)+\frac{i}{36}\bigg( \bar{\epsilon}^i\gamma_\mu\zeta_i\bigg)+\frac{i}{12}\varepsilon_{kl}\bigg(\bar{E}\bar{\epsilon}^k\psi^l_\mu
    \bigg)-\frac{i}{16}\bigg(\bar{\psi}_\mu^i\eta_i \bigg)+\text{h.c.}\\[5pt]
    \delta{C}_{\mu}&= \bar{\epsilon}^{i}\gamma_{\mu}\psi_{i}-2\bar{\epsilon}^{i}\psi_{\mu}^j \bar{\xi} \varepsilon_{ij} +\text{h.c.} \\[5pt]
    \delta\Tilde{C}_\mu&= -i\bar{\epsilon}^i\gamma_\mu\psi_i+2i\bar{\epsilon}^i\psi_\mu^j\varepsilon_{ij}\bar{\xi} +\text{h.c.} \\[5pt]
   \delta B_{\mu \nu} &= (\varepsilon_{ij} \bar{\epsilon}^i \gamma_{\mu \nu} \psi^j \bar{\xi} + 2 \xi \bar{\xi} \bar{\epsilon}^i \gamma_{[\mu} \psi_{\nu ]i}+\text{h.c} ) +\frac{1}{2}\left(C_{[\mu}\delta C_{\nu]}+\tilde{C}_{[\mu}\delta C_{\nu]}\right) \;,\label{Bmunu}\\[5pt]
    \delta E&=-\frac{1}{2}\varepsilon_{jk}\bar{\epsilon}^j\zeta^k+\frac{1}{2}\bar{\epsilon}^j \mathring{\chi}_j \\[5pt]
    \delta D^i{}_j&=-3\bar{\epsilon}^i \cancel{D} \zeta_j - 3 \varepsilon_{jk} \bar{\epsilon}^k \cancel{D} \mathring{\chi}^i  
     + \frac{1}{4} \epsilon_{jk} \bar{\epsilon}^i \zeta^k \bar{E}  
     + \frac{3}{4} \bar{\epsilon}^i \mathring{\chi}_j \bar{E} +\text{h.c.} -\text{trace}\\[5pt]
     \delta \zeta^i & = 3 \varepsilon^{ij} \cancel{D}E
\epsilon_j  + \varepsilon^{ij} \gamma \cdot \cancel{D} \mathring{T}^- \epsilon_j -4 \gamma \cdot R(V)^i{}_j \epsilon^j + 2i \gamma \cdot R(v) \epsilon^i  - 16 i \gamma \cdot R(A) \epsilon^i  \nonumber \\
&
- \frac{1}{2} D^i_j \epsilon^j + \frac{1}{4} \mathring{D} \epsilon^i  + \frac{3}{8} \bar{E} \gamma \cdot \mathring{T}^- \epsilon^i  
 - \varepsilon^{ij} \gamma \cdot \mathring{T}^- \eta_j  + 3 \varepsilon^{ij} E \eta_j  \\
\delta \xi & = \varepsilon_{jk} \bar{\epsilon}^j \psi^k \\
\delta \psi_i &= -\frac{1}{2} \gamma \cdot \mathcal{F}^+ \epsilon_i - 2 \varepsilon_{ij} \cancel{D} \bar{\xi} \epsilon^j -2\varepsilon_{ij} \bar{\xi} \eta^j  
\end{align}
\end{subequations}
\end{widetext}
This $32+32$ multiplet can be consistently decomposed into two parts: a $24+24$ multiplet and an $8+8$ vector multiplet. 
The $24+24$ sector can be identified with the $\mathcal{N}=2$ vector 
dilaton Weyl multiplet constructed in \cite{Butter:2017pbp}. 
To establish this identification, one must slightly adjust the conventions used in \cite{Butter:2017pbp} for the dilaton Weyl multiplet in the following way,
\begin{align}
        \mathcal{V}_\mu^i{}_j \;&\longrightarrow\; -\tfrac{1}{2}\,\mathcal{V}_\mu^i{}_j \,
\end{align}
\begin{align}  
        \delta_Q(\epsilon) \;\longrightarrow\; 
        \delta_Q(\epsilon) + \delta_K\!\left(
        \Lambda_{K\mu} = -\tfrac{1}{4}\,X^{-1}\,
        \bar{\epsilon}^i \gamma_\mu \slashed{D}\Omega_i + \text{h.c.}
        \right).
\end{align}
By making the above mentioned changes to the results of \cite{Butter:2017pbp}, one can see the mapping between the two multiplets as given below, 
where the components of the vector-dilaton Weyl multiplet are listed on the L.H.S and those of the truncated multiplet are listed on the R.H.S.

\allowdisplaybreaks
\begin{subequations}
\begin{align}
    e_{\mu}{}^{a} &=e_{\mu}{}^{a}   \\[5pt]
    \psi_{\mu}^{i} &= \psi_{\mu}^{i}   \\[5pt] 
  \mathcal{V}_{\mu}{}^{i}{}_{j}&=v_{\mu}{}^{i}{}_{j}\\[5pt]
    b_{\mu} &= b_{\mu}  \\[5pt]
  X&= \xi \\[5pt]
    W_\mu&=-C_\mu \\[5pt]
    \tilde{W}_\mu&=-\tilde{C}_\mu \\[5pt]
   \Omega_i &= \varepsilon_{ij} \, \psi^j \\[5pt]
   B_{\mu \nu} &= B_{\mu \nu}    
\end{align}
    
\end{subequations}
The composite gauge field $A_\mu$ corresponding to the $U(1)$ R symmetry in the vector dilaton Weyl multiplet (given in the L.H.S) also maps to the following combination of fields in the truncated multiplet (given in the R.H.S) that is also composite:
\begin{align}
A_\mu &= A_\mu +\mathring{v}_\mu
\end{align}
Now the remaining independent fields in the   truncated multiplet will combine in the following fashion (given in the R.H.S) to form the components of an $\NN=2$ vector multiplet (given in the L.H.S).
  \begin{align} \label{dict weyl-vector}
         W_\mu &= - \mathring{v}_\mu +2 A_\mu \nonumber\\
        X &=- \frac{i}{8}E \nonumber  \\
        \Omega_i &= -\frac{i}{16} \mathring{\chi}_{i}+ \frac{i}{16} \epsilon_{ij} \zeta^j \nonumber \\
        Y_{ij} &=\frac{i}{24} \varepsilon_{k(i}D^k{}_{j)}
    \end{align}
    To summarize, in this section we performed a particular supersymmetric truncation of the $SU(2)\times U(1)\times U(1)$ dilaton Weyl multiplet of $\NN=3$ conformal supergravity, that does not break any of the R-symmetries. The result of the truncation was a combination of the $\NN=2$ vector-dilaton Weyl multiplet and vector multiplet which are known in the literature.
\subsection{Supersymmetric Truncation-II}\label{susytrunc2}
In this subsection, we will perform another distinct truncation procedure, where we set one of the $\epsilon^i$ and the corresponding gravitino to zero. Without any loss of generality, we take it to be the second component as mentioned below.
\begin{align}
\epsilon^{2}&=0=\psi_{\mu}^{2}
\end{align}
Subsequent applications of supersymmetry transformations on the truncation condition mentioned above, we find that the following fundamental field components of the multiplet gets truncated out.
\begin{widetext}
    \begin{align}
T^{1}_{ab} = v_{\mu}{}^{2}{}_{1} = E^{1}= \zeta^{2}=\chi^{11}=\chi^{22}=\psi_R=\psi_{1}=Y_{a}^{2}=E= D^{2}{}_1= C_{\mu}=\tilde{C}_\mu=0
\end{align}
\end{widetext}
In addition, the $S$-supersymmetry parameter $\eta^2$ and the corresponding dependent gauge field $\phi_{\mu}^2$ also becomes zero. As a result, the following composite objects denoted with a mathring (see appendix-\ref{N3dilaton}) also becomes zero:
\begin{align}
\mathring{T}_{ab}=\mathring{\chi}^{1}=\mathring{\chi}=\mathring{D}^{2}=0
\end{align}
As expected, the $ SU(2)$ part of the $R$-symmetry gets broken to $U(1)$ and the $R$-symmetry realized on the truncated multiplet will be $U(1) \times U(1) \times U(1)$. The gauge fields associated with the three $U(1)$s are $-iv_\mu{}{}^2{}_2\equiv v_\mu$, $A_\mu$ and a composite $\mathring{v}_{\mu}$. We use $c_v$, $c_A$ and $c_{\mathring{v}}$ to denote the chiral weights of the fields w.r.t. the U(1) transformations associated with the gauge fields $v_\mu$, $A_\mu$ and $\mathring{v}_{\mu}$, respectively. Similarly, we will denote the U(1) gauge transformation parameters associated with the gauge fields $v_\mu$, $A_\mu$ and $\mathring{v}_{\mu}$ as $\lambda_v$, $\lambda_A$ and $\lambda_{\mathring{v}}$, respectively. For the sake of brevity, we rename the fundamental fields as well as composite objects that survive the truncation, as mentioned below:
\begin{widetext}
    \begin{align}
&\epsilon^1\to\tilde{\epsilon}_L \;,
\psi_{\mu}^1\to\tilde{\psi}_{\mu L}\;, 
\eta^1\to\tilde{\eta}_R \;,
\phi_{\mu}^1\to\tilde{\phi}_{\mu R}\;, 
-iv_\mu{}^2{}_2 \to {v}_\mu\;,
T_{ab}{}_{2} \to T^-_{ab}\;,
Y_a^1\to Y_a\;,
E_2 \to E\; \nonumber \\
& D^2{}_{2} \to {D}\;, 
\chi_{12}\to\chi_{L}\;,
\zeta^1\to \zeta_L\;,
\psi^2\to{\psi_L}\;,
\mathring{\chi}_2\to \mathring{{\chi}}_L\;,
\mathring{D}_1 \to \mathring{\widetilde{D}}
\end{align}
As a result of the above-mentioned truncations, we have a $32+32$ components multiplet which realises a $U(1) \times U(1)\times U(1)$ R-symmetry. The independent fields of the multiplet are tabulated in Table-\ref{new_multiplet_3} along with all its properties. 
\begin{table}[ht]
    \centering
    \begin{tabular}{|c|c|c|c|c|c|}
        \hline
        Field & Properties & $w$ & $c_A$ & $c_v$ & $c_{\mathring{v}}$ \\
        \hline
        \multicolumn{6}{|c|}{Independent Gauge fields}\\
        \hline
        $e_\mu^a$ & vielbein & $-1$ & 0 & 0 & 0\\
        $\tilde{\psi}_{\mu L}$ & $\gamma_5 \tilde{\psi}_{\mu}=\tilde{\psi}_{\mu L}$ & $-1/2$ & $-1/2$ & $-1$ & $-1/2$\\
        $\psi_{\mu L}$ & $\gamma_5 \psi_\mu=\psi_{\mu L}$ & $-1/2$ & $-1/2$ & 0 & 1\\
        $v_\mu$ & $U(1)$ gauge field& 0 & 0 & 0 & 0\\
        $A_\mu$ & $U(1)_A$ gauge field & 0 & 0 & 0 & 0\\
        $b_\mu$ & Dilatation gauge field & 0 & 0 & 0 & 0\\
        $B_{\mu \nu}$ & Two-form gauge field & 0 & 0 & 0 & 0\\
        \hline
        \multicolumn{6}{|c|}{Covariant fields}\\
        \hline
        $Y_a$ & Boson & 1 & $0$ & $-1$ & $-3/2$\\
        $T_{ab}^-$ & Anti-self dual& 1 & $-1$ & $-1$ &$1/2$\\
        $E$ & Complex scalar & 1 & $-1$ & $-1$ & $1/2$\\
        $D$ & Real scalar & 2 & 0 & 0 & 0\\
        $\chi$ & $\gamma_5\chi =\chi_L $ & $3/2$ & $-1/2$ & $0$ & 1\\
        $\zeta_L$ & $\gamma_5\zeta=\zeta_L$ & $3/2$ & $-1/2$ & $-1$ & $-1/2$\\
        $\xi$ & Complex scalar & 1 & $-1$ & 0 & $-1$\\
        $\psi_R$ & $\gamma_5 \psi = - \gamma_5\psi_R$ & $3/2$ & $1/2$ & $-1$ & $1/2$\\
        $\theta_L$ & $\gamma_5 \theta_L = \theta_L$ & $3/2$ & $3/2$ & 0 & 0\\
        \hline	
    \end{tabular}
     \caption{Field content of the new $\mathcal{N}=2$ multiplet with R-symmetry $U(1)\times U(1)\times U(1)$}
		\label{new_multiplet_3}
    \end{table}

    The supersymmetry transformations of the components are as given below:
 \allowdisplaybreaks
{\begin{subequations}\label{2dilatontrans}
\begin{align}
\delta e_\mu^a&=\bar{\tilde{\epsilon}}_R\gamma^a\tilde{\psi}_{\mu L}+\bar{\epsilon}_R\gamma^a\psi_{\mu L}+ \text{h.c.} \\[5pt]
\delta\tilde{\psi}_{\mu L}&=2\mathcal{D}_\mu\tilde{\epsilon}_L    -2Y_\mu\epsilon_L-u({\psi}_{\mu})\epsilon_L-\frac{1}{8} \gamma\cdot T^-\gamma_\mu\epsilon_R +u({\epsilon})\psi_{\mu L}-\gamma_\mu \tilde{\eta}_R \\[5pt]
\delta \psi_{\mu L}&= 2\mathcal{D}_\mu\epsilon_L +2\bar{Y}_{\mu} \tilde{\epsilon}_L + \bar{u}({\psi}_{\mu}) \tilde{\epsilon}_L +\frac{1}{8}\gamma\cdot T^-\gamma_\mu \tilde{\epsilon}_R -\bar{u}({\epsilon})\tilde{\psi}_{\mu L}  -\gamma_\mu\eta_R\\[5pt]
    \delta v_\mu& =\frac{i}{2}\bar{\tilde{\epsilon}} \tilde{\phi}_{\mu L}-\frac{i}{2}\bar{\tilde{\psi}}_{\mu L}\tilde{\eta}_{L}-\frac{i}{96}\bar{\tilde{\epsilon}}_{L}\gamma_\mu\zeta_R+\frac{i}{32}\bigg(\bar{\tilde{\epsilon}}_L\gamma_\mu \mathring{\chi}_{R}-\bar{\epsilon}_L\gamma_\mu\chi_{R} \bigg)+\frac{i}{16}\bar{E}\bigg(\bar{\tilde{\epsilon}}_L \psi_{\mu L}-\bar{\epsilon}_L\tilde{\psi}_{\mu L} \bigg)  \nonumber \\
    &+
    \frac{1}{2}u(\tilde{\epsilon}_L)\bar{Y}_{\mu} +\frac{1}{2}u(\tilde{\epsilon}_L)u(\tilde{\psi}_{\mu R})
     +\text{h.c} \\[5pt]
   \delta b_\mu&=\frac{1}{2}\bigg(\bar{\tilde{\epsilon}}_L\tilde{\phi}_{\mu L}+\bar{\epsilon}_L\phi_{\mu L}-\bar{\tilde{\psi}}_{\mu L} \tilde{\eta}_L-\bar{\psi}_{\mu L}\eta_L \bigg)+\text{h.c.} \\[5pt]
    \delta A_\mu&= \frac{i}{6}\bigg( \bar{\tilde{\epsilon}}_L\tilde{\phi}_{\mu L}+\bar{\epsilon}_L\phi_{\mu L}\bigg)+\frac{i}{36}\bigg(\bar{\tilde{\epsilon}}_L \gamma_\mu\zeta_R +\bar{\epsilon}_L\gamma_\mu \mathring{\zeta}_R\bigg)+\frac{i}{12}\bigg(-\bar{E}\bar{\tilde{\epsilon}}_L\psi_{\mu L}
    +\bar{E}\bar{\epsilon}_L\tilde{\psi}_{\mu L}
    \bigg) \nonumber \\
    & -\frac{i}{16}\bigg(\bar{\tilde{\psi}}_{\mu L} \tilde{\eta}_L +\psi_{\mu L}\eta_L \bigg)+\text{h.c.}\\[5pt]     
      \delta B_{\mu \nu} &= (\bar{\epsilon}_R \gamma_{\mu \nu} \theta_R \bar{\xi} + \bar{\tilde{\epsilon}}_L \gamma_{\mu \nu} \psi_L \bar{\xi}  - 2\bar{\xi} \xi \bar{\epsilon}_R \gamma_{[\mu} \psi_{\nu] L} + 2 \bar{\xi}\xi \bar{\tilde{\epsilon}}_L \gamma_{[\mu} \tilde{\psi}_{\nu ]R} + \text{h.c})  + 2 \partial_{[\mu} \Lambda_{\nu ]}\\[5pt]
       \delta Y_a& = \bar{\tilde{\epsilon}}_L D_a(\frac{\theta_L}{{\bar{\xi}}}) + \bar{\epsilon}_R D_a(\frac{\psi_R}{\bar{\xi}}) -\frac{1}{48} \bar{\tilde{\epsilon}}_L \gamma_a \mathring{\zeta}_R + \frac{1}{48} \bar{\epsilon}_R \gamma_a \zeta_L + \frac{1}{16}  \bar{\tilde{\epsilon}}_L \gamma_a \chi_{R} \nonumber  + \frac{1}{16}  \bar{\epsilon}_R \gamma_a \mathring{\chi}_{L} \\& - \frac{1}{2 {\bar{\xi}}} \bar{\theta}_L \bigg( -2Y_a\epsilon_L -\frac{1}{8}  \gamma \cdot T^- \gamma_a \epsilon_R \nonumber \bigg) -\frac{1}{{\bar{\xi}}}  \bar{\psi}_R \tilde{\epsilon}_R Y_a + \frac{1}{2 \bar{\xi}} \bar{\theta}_L \gamma_a \tilde{\eta}_L
    + \frac{1}{2\bar{\xi}}  \bar{\psi}_R \gamma_a \eta_L \\[5pt] 
    \delta E&= \frac{1}{2}\bigg(\bar{\tilde{\epsilon}}_L
    \mathring{\zeta}^L-\bar{\epsilon}_L\zeta_L \bigg)+\frac{1}{2}\bar{\tilde{\epsilon}}_L
    \chi_{L}+\frac{1}{2}\bar{\epsilon}_L \mathring{\chi}_{L} \\[5pt]    
    \delta {T}^+_{ab} &= 4 \bar{\tilde{\epsilon}}_R R_{ab}(Q)_R - 4 \bar{\epsilon}_R \tilde{R}_{ab}(Q)_{R} + \frac{1}{8} \bar{\tilde{\epsilon}}_R
    \gamma_{ab} \chi_{R} + \frac{1}{8} \bar{\epsilon}_R \gamma_{ab} \mathring{{\chi}}_{R} \nonumber \\
    &- \frac{1}{24} \bar{\tilde{\epsilon}}_R \gamma _{ab} \mathring{\zeta}_R + \frac{1}{24} \bar{\epsilon}_R \gamma_{ab} \zeta_R  \\[5pt]        
   \delta D&= \frac{3}{2}\bar{\tilde{\epsilon}}_L \cancel{D} \mathring{{\chi}}_{R}+ \frac{3}{2}\bar{\tilde{\epsilon}}_L \slashed{\bar{Y}} \chi_{R} - \frac{3}{2}\bar{\epsilon}_L \cancel{D} \chi_{R}   + \frac{3}{2} \bar{\epsilon}_L \slashed{\bar{Y}} \mathring{{\chi}}_{R} - \frac{1}{8} \bar{\tilde{\epsilon}}_L \mathring{\zeta}_L \bar{E} +\frac{1}{4} \bar{\epsilon}_L \zeta_L \bar{E}\nonumber \\
    & -\frac{1}{2}\bar{\tilde{\epsilon}}_L \cancel{D} \zeta_{R} +\frac{3}{2}\bar{\tilde{\epsilon}}_L \slashed{\bar{Y}} \mathring{\zeta}_{R}+\frac{1}{2}\bar{{\epsilon}}_L \slashed{Y} \mathring{\chi}_{R} -\frac{3}{8} \bar{\tilde{\epsilon}}_L \chi_L \bar{E} + \text{h.c.}  \\[5pt]
    \delta \chi_{L} &= \cancel{D}E \tilde{\epsilon}_{R} - 8 \varepsilon_{2(1} \gamma \cdot R(V)^{2}{}_{2)} \epsilon_L - 8 \gamma ^{ab} D_a \bar{Y}_{b} \tilde{\epsilon}_{L}  + \frac{2}{ \xi}  \gamma ^{ab} \bar{\theta}_R \tilde{R}(Q)_{ab R} \tilde{\epsilon}_{L}   \nonumber \\
    & + \frac{2}{\xi} \gamma^{ab} \bigg( \bar{{\psi}}_L R(Q)_{ab L} \bigg) \tilde{\epsilon}_{L}  
    -  \gamma \cdot \cancel{D} T^- \tilde{\epsilon}_{R}  -\frac{1}{6} \mathring{\tilde{D}} \tilde{\epsilon}_{L}  +\frac{1}{8}  \bar{E} \gamma \cdot T^- \epsilon_L  - \frac{1}{8} E \bar{E}\epsilon_L + \nonumber\\
    & -\frac{1}{3} {D} \epsilon_{L} + u(\tilde{\epsilon}_{R}) \mathring{{\chi}}_{L}  +  \gamma \cdot T^- \tilde{\eta}_{L} + E \tilde{\eta}_L  \\  
     \delta \zeta_L & =- 3 \cancel{D} E \epsilon_R  -  \gamma \cdot \cancel{D}T^- \epsilon_R - 4 \gamma^{ab} Y_a \bar{Y}_{b} \tilde{\epsilon}_{L}   - 18 i \gamma \cdot R(A) \tilde{\epsilon}_{L}   + \frac{1}{4} \mathring{D} \tilde{\epsilon}_{L}   \nonumber \\
& -\frac{1}{2} \mathring{\bar{\tilde{D}}} \epsilon_L  + \frac{3}{8} \bar{E} \gamma \cdot T^- \tilde{\epsilon}_{L}  + \frac{1}{2} {D} \tilde{\epsilon}_{L}   + \frac{1}{8} \bar{E} E \tilde{\epsilon}_{L}   + u(\tilde{\epsilon}_{L}) \mathring{\zeta}_L + \gamma \cdot T^- \eta_L - 3 E \eta_L  \\[5pt]    
\delta \xi & = -   \bar{\epsilon}_R \theta_R +  \bar{\tilde{\epsilon}}_L {\psi}_L \\[5pt]
\delta {\psi}_R &= 2  \cancel{D} \bar{\xi} \tilde{\epsilon}_{L}   + 2 \slashed{Y} \bar{\xi} \epsilon_L  -\frac{1}{4} E \bar{\xi} \epsilon_R + 2 \bar{\xi} \tilde{\eta}_R \\[5pt] 
\delta \theta_L & = 2 \slashed{Y} \bar{\xi} \tilde{\epsilon}_R - 2 \cancel{D}\bar{\xi} \epsilon_R +\frac{1}{4} \bar{E} \bar{\xi} \tilde{\epsilon}_{L}   - 2 \bar{\xi} \eta_L 
\end{align}
    \end{subequations}}
The expressions for the composite objects appearing in the above-mentioned transformations are as follows:
\begin{subequations}\label{fermsol}
\begin{align}
u(\epsilon)&=\frac{1}{\bar{\xi}}\bigg(\bar{\tilde{\epsilon}}_L \theta_L+\bar{\epsilon}_R\psi_R\bigg) \\
 \mathring{v}_a =& \frac{i}{2\xi\bar{\xi}} \bigg(- \frac{1}{3!} \varepsilon_{abcd} H^{bcd} + \bar{\xi}\partial_a\xi-\xi\partial_a\bar{\xi}+2iA_a\bar{\xi}\xi +(\frac{1}{2}\bar{\xi}\bar{\psi}_{a,R}\theta_R-\frac{1}{2}\bar{\xi}\bar{\tilde{\psi}}_{a}{}_{L}\psi_L -h.c ) \nonumber\\
 &+\frac{1}{2}\bar{\theta}_R\gamma_a\theta_L+\frac{1}{2}\bar{\psi}_L\gamma_a\psi_R \bigg)
\\
     \mathring{{\chi}}_{L}  & = -\frac{8}{ \bar{\xi}} \bigg(  \cancel{D} {\psi}_R -\frac{1}{8} E \theta_L+\frac{1}{8} \gamma \cdot T^- \theta_L-\frac{1}{24} \zeta_L\bar{\xi} \bigg)\;,\label{chisol} \\     
    \mathring{\zeta}_L  & = \frac{12}{\xi}  \bigg(  \cancel{D} \theta_R -\frac{1}{8} E {\psi}_L -\frac{1}{8} \gamma \cdot T^- {\psi}_L \bigg)\;.\label{zeta3sol}
\\
     \mathring{\widetilde{D}} &= \frac{48}{\xi} \bigg( - D^a(\bar{Y}_{a} \xi) -\bar{Y}_{a} D_a \xi  -\frac{1}{16} \bar{\chi}_{L} {\psi}_L -\frac{1}{48} \mathring{\bar{\zeta}}_L {\psi}_L \bigg) \;\label{Disol1}\\
 \mathring{D}&=\frac{24}{{\lvert\xi \rvert}^2 } \bar{\xi}\bigg\{D^aD_a \xi + Y^{a}\bar{Y}_{a} \xi +\frac{1}{24} \mathring{\bar{\zeta}}_R \theta_R -\frac{1}{16} \mathring{\bar{{\chi}}}_{L} {\psi}_L -\frac{1}{48}  \bar{\zeta}_L {\psi}_L -\frac{1}{96} \xi \bar{E} E +\text{h.c.}\bigg\}\;\label{Dsol}  
\end{align}
\end{subequations}  
         \end{widetext}
    To summarize, in this section we performed a particular supersymmetric truncation on the $SU(2)\times U(1) \times U(1)$ dilaton Weyl multiplet of $\NN=3$ conformal supergravity \cite{Adhikari:2024qxg} and arrived at an $\NN=2$ multiplet which has $32+32$ components and realizes a $U(1)\times U(1)\times U(1)$ R-symmetry. 
\section{Construction of a new $\mathcal{N}$=2 multiplet with $U(1) \times U(1) \times U(1)$ R-symmtery}\label{N2const}
$\NN=2$ conformal supergravity has been extensively studied in the literature. The construction of the standard Weyl multiplet was first presented in \cite{Bergshoeff:1980is}, and several matter multiplets for $\NN=2$ conformal supergravity are now well established. The formulation of their actions has been made possible through various well-known density formulas, such as the chiral density formula \cite{deRoo:1980mm}, the linear-vector density formula \cite{deWit:1982na}, and a newer approach based on an abstract fermionic multiplet \cite{Hegde:2019ioy}. Given the vastness of the literature on $\NN=2$ conformal supergravity, we refrain from citing all related works for brevity.

Recently, a new matter multiplet, referred to as the scalar-tensor multiplet, was constructed within the $\NN=2$ conformal supergravity framework \cite{Aikot:2024cne}. This multiplet arises from an on-shell massive $\NN=2$ hypermultiplet with a broken rigid $SU(2)$ symmetry, charged under the $U(1)$ gauge symmetry of an off-shell $\NN=2$ vector multiplet (referred to as the central charge $U(1)$). By using the field equations of the hypermultiplet, one can algebraically eliminate certain fields from the vector multiplet, while introducing a 2-form gauge field.\footnote{The procedure involved in this construction is very similar to the construction of dilaton Weyl multiplets and the details can be found in \cite{Aikot:2024cne}.} The resulting multiplet carries $8+8$ off-shell degrees of freedom and is identified as the scalar-tensor multiplet.

Further details regarding the  $\NN=2$ standard Weyl multiplet and the scalar-tensor multiplet can be found in Appendix-\ref{N2review}.
 
The $SU(2)$ R-symmetry group of the $\NN=2$ theory is broken to $U(1)$ by imposing the following condition on a scalar field belonging to the scalar-tensor multiplet: 
\begin{subequations}
\begin{align}
 \xi^2&=0
\end{align}    
\end{subequations}
Since, this leaves us with a single component $\xi_1$, we will henceforth rename it as $\xi$. The Q-supersymmetry transformation does
not preserve the gauge condition. Hence, it needs to be redefined in the following way by adding a compensating $SU(2)$ transformation;
\begin{align}
    \delta^{new}_{Q}(\epsilon) &= \delta_{Q}(\epsilon) + \delta_{SU(2)}(\Lambda^2{}_1 =u(\epsilon))
\end{align}
Where,
\begin{subequations}
\begin{align}
u(\epsilon) &= \frac{1}{\bar{\xi}}\left(\bar{\epsilon}^2 \theta_{L}+\bar{\epsilon}_{1}\psi_{R}\right)\\
\end{align}    
\end{subequations}
The $U(1)\subset SU(2)$ that survives the breaking combines with the other $U(1)$ R-symmetry of the $\NN=2$ theory and the $U(1)$ central charge transformation present in the scalar-tensor multiplet to give rise to the symmetry group $U(1)\times U(1)\times U(1)$ realized on the new multiplet. The gauge fields corresponding to the first two $U(1)$'s are independent gauge fields: $-iV_\mu{}^2{}_2\equiv v_\mu$ and $A_\mu$ respectively whereas the gauge field corresponding to the last $U(1)$ is the composite gauge field $\mathring{W}_{\mu}$. We use $c_v$, $c_A$ and $c_{z}$ to denote the chiral weights of the fields w.r.t. the $U(1)$ transformations associated with the gauge fields $v_\mu$ and $A_\mu$ and $\mathring{W}_{\mu}$ respectively. In table \ref{new_multiplet}, we give the values of $c_v$, $c_A$ and $c_{z}$ for all the fields of the multiplet. Similarly, we will denote the $U(1)$ gauge transformation parameters associated
with the gauge fields $v_\mu$ $A_\mu$ and $\mathring{W}_{\mu}$ as $\lambda_v$, $\lambda_A$ and $\lambda_{z}$ respectively.
The component $ {V}_{a}{}^2{}_1$  will turn into covariant field after making appropriate modification to it by adding gravitino dependent terms as shown below. 
\begin{align}
    Y_a &=  V_{a}{}^2{}_1 - \frac{1}{2}u(\psi_{a})
\end{align} 
After completing the above-mentioned exercise, the components of the $\NN=2$ standard Weyl multiplet and the scalar-tensor multiplet fuse together to give rise to a $32+32$ component multiplet tabulated in Table-\ref{new_multiplet}.
\begin{widetext}
We give the supersymmetry transformations of the fields of this multiplet below:
    \begin{table}[ht]
    \centering
    \begin{tabular}{|c|c|c|c|c|c|}
        \hline
        Field & Properties & $w$ & $c_A$ & $c_v$ & $c_{z}$ \\
        \hline
        \multicolumn{6}{|c|}{Independent Gauge fields}\\
        \hline
        $e_\mu^a$ & vielbein & $-1$ & 0 & 0 & 0\\
        ${\psi}_{\mu}^1$ & $\gamma_5 {\psi}^1_{\mu}={\psi}_{\mu}^1$ & $-1/2$ & $-1/2$ & $-1$ & $0$\\
        $\psi_{\mu}^2$ & $\gamma_5 \psi_\mu^2=\psi_{\mu}^2$ & $-1/2$ & $-1/2$ & 1 & $0$\\
        $v_\mu$ & $U(1)$ gauge field& 0 & 0 & 0 & 0\\
        $A_\mu$ & $U(1)_A$ gauge field & 0 & 0 & 0 & 0\\
        $b_\mu$ & Dilatation gauge field & 0 & 0 & 0 & 0\\
        $B_{\mu \nu}$ & Two-form gauge field & 0 & 0 & 0 & 0\\
        \hline
        \multicolumn{6}{|c|}{Covariant fields}\\
        \hline
        $Y_a$ & Boson & 1 & $0$ & $2$ & $0$\\
        $T_{ab}^{12}$ & Anti-self dual& 1 & $-1$ & 0 &$0$\\
        $X$ & Complex scalar & 1 & $-1$ & $0$ & $0$\\
        $D$ & Real scalar & 2 & 0 & 0 & 0\\
        $\chi^1$ & $\gamma_5\chi^1 =\chi^1 $ & $3/2$ & $-1/2$ & $-1$ & 0\\
        $\chi^2$ & $\gamma_5\chi^2=\chi^2$ & $3/2$ & $-1/2$ & $1$ & $0$\\
        $\xi$ & Complex scalar & 1 & $0$ & 1 & $-1/2$\\
        $\psi_R$ & $\gamma_5 \psi = - \gamma_5\psi_R$ & $3/2$ & $-1/2$ & $0$ & $1/2$\\
        $\theta_L$ & $\gamma_5 \theta_L = \theta_L$ & $3/2$ & $1/2$ & 0 & $1/2$\\
        \hline	
    \end{tabular}
     \caption{Field content of the $\mathcal{N}=2$  multiplet with R-symmetry $U(1)\times U(1)\times U(1)$} after fusing with Scalar-Tensor multiplet
		\label{new_multiplet}
    \end{table}
    \begin{subequations}
\begin{align}\label{WeylTransf}
\delta e_{\mu}{}^{a} &= \bar{\epsilon}_{1}\gamma^{a}\psi^1_{\mu}+ \bar{\epsilon}_{2}\gamma^{a}\psi^2_{\mu}+\text{h.c.}\\[5pt]
\delta\psi_\mu^1&=2\mathcal{D}_\mu\epsilon^1  +2\bar{Y}_\mu\epsilon^2+\bar{u}(\psi_\mu)\epsilon^2-\frac{1}{8} \gamma\cdot T^{12}\gamma_\mu\epsilon_2 -\bar{u}(\epsilon)\psi_{\mu }^2-\gamma_\mu \eta^1  \\[5pt]
\delta \psi_{\mu}^2&= 2\mathcal{D}_\mu\epsilon^2 -2Y_{\mu}\epsilon^1 - u(\psi_\mu)\epsilon^1 +\frac{1}{8}\gamma\cdot T^{12}\gamma_\mu \epsilon_1+ u(\epsilon)\psi^1_\mu  -\gamma_\mu\eta^2  \\[5pt]
\delta v_{\mu}&=\frac{i}{2}\bar{\epsilon}^{1}\phi_{\mu 1} -\frac{i}{2}\bar{\epsilon}^{2}\phi_{\mu 2} -\frac{i}{2}\bar{\epsilon}^{1}\gamma_{\mu}\chi_{1}+\frac{i}{2}\bar{\epsilon}^{2}\gamma_{\mu}\chi_{2} +\frac{i}{2}\bar{\eta}^{1}\psi_{\mu 1}-\frac{i}{2}\bar{\eta}^{2}\psi_{\mu 2}+\text{h.c.}   \\[5pt]
\delta b_{\mu}&=\frac{1}{2}\bar{\epsilon}^{1}\phi_{\mu 1} +\frac{1}{2}\bar{\epsilon}^{2}\phi_{\mu 2} -\frac{1}{2}\bar{\eta}^{1}\psi_{\mu 1}-\frac{1}{2}\bar{\eta}^{2}\psi_{\mu 2}+\text{h.c.}    \\[5pt]
\delta A_{\mu} &= \frac{i}{2}\bar{\epsilon}^{1}\phi_{\mu 1}+ \frac{i}{2}\bar{\epsilon}^{2}\phi_{\mu 2}+ i\bar{\epsilon}^{1}\gamma_{\mu}\chi_{1}+i\bar{\epsilon}^{2}\gamma_{\mu}\chi_{2}+ \frac{i}{2}\bar{\eta}^{1}\psi_{\mu 1} +\frac{i}{2}\bar{\eta}^{2}\psi_{\mu 2} + \text{h.c.}   \\[5pt]
\delta T_{ab}{}^{12}&=  8\,\bar{\epsilon}^{[1}R(Q)_{ab}{}^{2]}-4\bar{\epsilon}^{[1}\gamma_{ab}\chi^{2]}  \\[5pt]
\delta\chi^{1}&=-\frac{1}{12}\gamma \cdot T^{12}\epsilon_{2} + \frac{1}{6} \gamma \cdot R(V)^{1}{}_{2}\epsilon^{2} - \frac{i}{3}\gamma \cdot R(A)\epsilon^{1}+D\epsilon^{1}+\frac{1}{12}\gamma\cdot T^{12}\eta_{2}  \\[5pt]
\delta\chi^{2}&=-\frac{1}{12}\gamma \cdot T^{21}\epsilon_{1} + \frac{1}{6} \gamma \cdot R(V)^{2}{}_{1}\epsilon^{1} - \frac{i}{3}\gamma \cdot R(A)\epsilon^{2}+D\epsilon^{2}+\frac{1}{12}\gamma\cdot T^{21}\eta_{1}  \\[5pt]
\delta D&= \bar{\epsilon}^{1}\cancel{D}\chi_{1} + \bar{\epsilon}^{2}\cancel{D}\chi_{2} +\text{h.c}\;\\[5pt]
\delta \psi_{R}&=  2\cancel{D} \bar{\xi} +2\slashed{Y} \bar{\xi} {\epsilon}^1 - 2i X \bar{\xi} \epsilon_1 + 2  \bar{\xi} {\eta}^2   \\[5pt]
\delta \theta_L &= - 2 \cancel{D} \bar{\xi} \epsilon_1 +2\slashed{Y} \bar{\xi} {\epsilon}_2 - 2i\bar{X} \bar{\xi} \epsilon^2  - 2 \bar{\xi} \eta_1   \\[5pt]
\delta \xi&=-\bar{\epsilon_1} \theta_R +  \bar{\epsilon}^2 \psi_L   \\[5pt]
\delta B_{\mu \nu} & = \bar{\epsilon}_1 \gamma_{\mu \nu} \theta_R \bar{\xi} + \bar{\epsilon}^2 \gamma_{\mu \nu} \psi_L \bar{\xi} -2 \xi \bar{\xi} \bar{\epsilon}_1 \gamma_{[\mu} \psi_{\nu]}^1 + 2 \bar{\xi} \xi \bar{\epsilon}_2 \gamma_{[\mu} \psi_{\nu]}^2 + \text{h.c}  
\end{align}
\end{subequations}
\end{widetext}
\section{Relating the two multiplets}\label{Relation}
In subsection-\ref{susytrunc2}, we performed a particular supersymmetric truncation on the $\NN=3$ dilaton Weyl multiplet that broke the $SU(2)\times U(1)\times U(1)$ R-symmetry to $U(1)\times U(1)\times U(1)$. The result was an $\NN=2$ multiplet which has $32+32$ degrees of freedom. In section-\ref{N2const}, we constructed a new $\NN=2$ multiplet which fused a scalar-tensor multiplet with the $\NN=2$ standard Weyl multiplet by using a scalar field of the scalar-tensor multiplet to break the $SU(2)$ part of the R-symmetry to $U(1)$. The resulting $\NN=2$ multiplet also has $32+32$ degrees of freedom. Let us make the following identifications, where on the LHS we have the fields from the new $\mathcal{N}=2$ multiplet and on the RHS we have the fields arising from the supersymmetric truncation of the $\NN=3$ dilaton Weyl multiplet.
\allowdisplaybreaks
\begin{subequations}
    \begin{align}
    e_{\mu}{}^{a} &=e_{\mu}{}^{a}   \\[5pt]
    \psi_{\mu}^{1} &= \psi_{\mu L}   \\[5pt]
    \psi_{\mu}^{2} &= \tilde{\psi}_{\mu L}   \\[5pt]
     T_{ab}^{12} {}{} &= -T^-_{ab}     \\[5pt]
 v_{\mu} &= \frac{i}{2}v_{\mu}{}^2{}_2 -\frac{3}{4}\mathring{v}_{\mu}  \\[5pt]
    b_{\mu} &= b_{\mu}  \\[5pt]
    A_{\mu} &= A_{\mu} - i\left({v} _{\mu}{}^{2}{}_{2}-\frac{i}{2} \mathring{v}_{\mu}\right)   \\[5pt]
  Y_{a}&=Y_{a}
  \\[5pt]
    D &= \frac{1}{48}\left({D} +\frac{1}{2}\mathring{D}+\frac{1}{2} |E|^2\right)    \\[5pt]
   \chi^{1} &= \frac{-1}{16}\left(-\frac{1}{3}\mathring{\zeta}_{L} +\chi_{L}\right)        \\[5pt]
    \chi^{2} &= \frac{1}{16}\left(\frac{1}{3}\zeta_R +\mathring{{\chi}}_{R}\right)   \\[5pt]
  \xi&= \xi \\[5pt]
   \psi_R &= \psi_R \\[5pt]
   \theta_L &= \theta_L \\[5pt]
   B_{\mu \nu} &= B_{\mu \nu} \\[5pt]
   X &= \frac{i}{8}E 
\end{align}
\end{subequations}
The above mentioned identifications of the independent fields on both sides results in the following identifications of the composite fields on both sides. 
\begin{align}
    \mathring{W}_{a}&=2A_{a}+iv_{a}{}^2{}_2-\frac{1}{2}\mathring{v}_{a}
\end{align}
It is a straightforward exercise to see that the above-mentioned identifications lead to the same supersymmetry transformation rules of both the multiplets thereby establishing the equivalence of both the multiplets. 
\section{Discussions}
Supersymmetry truncation procedure relates multiplets in conformal supergravity theories with higher supersymmetries to multiplets with lower supersymmetries. For example, $\NN=3$ standard Weyl multiplet is related to $\NN=4$ standard Weyl multiplet \cite{Hegde:2021rte}, $\NN=2$ standard Weyl multiplet is related to $\NN=3$ standard Weyl multiplet \cite{Aikot:2024cne} and $\NN=1$ standard Weyl multiplet is related to $\NN=2$ standard Weyl multiplet \cite{Yamada:2019ttz} via appropriate supersymmetric truncation procedure. We also know that the Weyl multiplet can appear in two variants: standard and dilaton, in conformal supergravity theories. Hence, it is natural to explore the supersymmetric truncation procedure on dilaton Weyl multiplets similar to what has been done for the standard Weyl multiplets in the literature. 

As a case study, we implement the supersymmetric truncation procedure on the $\NN=3$ dilaton Weyl multiplet which realizes an $SU(2)\times U(1) \times U(1)$ R-symmetry \cite{Adhikari:2025wwb}. This truncation procedure can be carried out in two inequivalent ways. One of the truncation procedure does not break any of the R-symmetries of the $\NN=3$ dilaton Weyl multiplet and results in an $\NN=2$ vector-dilaton Weyl multiplet and the vector multiplet which we have shown explicitly in subsection-\ref{susytrunc1}. The second truncation procedure breaks the $SU(2)$ part of the R-symmetry to $U(1)$. This results in a gravity multiplet of the $\NN=2$ theory that realizes an $U(1)\times U(1)\times U(1)$ local symmetry instead of the $SU(2)\times U(1)$ R-symmetry and has $32+32$ degrees of freedom as shown in subsection-\ref{susytrunc2}. Such a multiplet is not known in the literature for $\NN=2$ conformal supergravity. The only two Weyl multiplets known for $\NN=2$ conformal supergravity are the standard Weyl multiplet \cite{deWit:1979dzm} and the dilaton Weyl multiplet \cite{Butter:2017pbp} which carries $24+24$ components and realizes an $SU(2)\times U(1)$ R-symmetry. In order to gain a better understanding of the multiplet arising from the second truncation procedure, we perform an independent construction in $\NN=2$ conformal supergravity in section-\ref{N2const}. In this construction, we couple an $\NN=2$ scalar-tensor multiplet \cite{Aikot:2024cne} with the standard Weyl multiplet and use one of the scalars of the scalar tensor multiplet to break the $SU(2)$ R-symmetry to $U(1)$. As a result, the R-symmetry becomes $U(1)\times U(1)$. This combines with an already existing $U(1)$ gauge symmetry in the scalar tensor multiplet, corresponding to the ``central-charge transformation'' to give us a $U(1)\times U(1)\times U(1)$ local symmetry which is also what was obtained using the supersymmetry truncation procedure. The resulting multiplet carries $32+32$ components which is equal to the components of the multiplet arising from the supersymmetric truncation procedure. The gauge field corresponding to the central-charge $U(1)$ originating from the scalar tensor multiplet is a dependent gauge field which confirms with the gauge field of one of the $U(1)$ coming from the supersymmetric truncation of the $\NN=3$ dilaton Weyl multiplet. These suggests that the multiplet arising from the supersymmetric truncation and the multiplet independently constructed in the $\NN=2$ theory are equivalent and in fact we explicitly show the equivalence between the two multiplets in section-\ref{Relation}. 

The resulting multiplet in subsection-\ref{susytrunc2} and section-\ref{N2const} is in fact gauge equivalent to a Poincar{\'e} supergravity multiplet since it has all the components necessary to compensate for the extra symmetries: the complex scalar $X$ to compensate for dilatation and $U(1)_A$, the imaginary part of the complex scalar $\xi$ to compensate for $U(1)_v$, the Majorana spinors $\theta$ and $\psi$ to compensate for $S$-supersymmetry and the gauge field $b_\mu$ to compensate for the special conformal transformation. This aspect of the supersymmetric truncation of the $\NN=3$ dilaton Weyl multiplet i.e giving rise to a gravity multiplet gauge equivalent to the Poincar{\'e} supergravity multiplet instead of a Weyl multiplet is unique and has not been observed before in any of the supersymmetric truncation procedures implemented on the standard Weyl multiplet. This motivates us to study the supersymmetric truncation procedure on the other known dilaton Weyl multiplets in both $\NN=4$ as well as $\NN=3$ theories \cite{Ciceri:2024xxf, Adhikari:2024esl, Adhikari:2025wwb}. This is currently a work in progress and we hope to report on some of the results in due course. 

We would also like to use the gravity multiplet constructed in this paper in subsection-\ref{susytrunc2} to construct a pure $\NN=2$ Poincar{\'e} supergravity theory. In this construction, we would not need any extra compensating multiplet since the gravity multiplet is already gauge equivalent to a Poincar{\'e} supergravity multiplet. Hence, the construction will be completely off-shell in nature. We can couple additional vector multiplets to this gravity multiplet and obtain matter coupled $\NN=2$ supergravity theory. One clear distinction between this construction and the earlier constructions of $\NN=2$ Poincar{\'e} supergravity is that the graviphoton, which is a gauge field associated with the central charge transformation, is a dependent gauge field. It would be nice to see the relation of such a construction with the earlier known constructions and use it to study black holes originating in the $\NN=2$ compactifications of string theory. We leave to address these questions in future.
\acknowledgments
SA and AA thanks Chennai Mathematical Institute for its hospitality during the course of this work. BS would like to thank the organizers of the conference ``Fields and Strings 2025'' held at the Steklov Mathematical Institute, Moscow where this work was presented.

\appendix
\section{Notations and Convention}\label{Notation}
We are working in four dimensions where the spinors are Majorana. We follow a chiral notation where the placement of the $SU(N)$ R-symmetry indices $i,j\cdots$ on a field typically denotes a particular projection of that field. For example, on a spinor such as the gravitino, placing the R-symmetry index on the superscript $\psi_{\mu}^{i}$ gives a left chiral projection and is related to the right chiral projection $\psi_{\mu i}$ via complex conjugation as a result of the Majorana condition, as shown below:
\begin{align}
    \psi_{\mu i}=i\gamma^0C^{-1}\left(\psi_{\mu}^{i}\right)^{*}\;,
\end{align}
where $C$ is the charge conjugation Matrix. Similarly the positioning of the indices on an anti-symmetric real tensor field gives a particular duality projection of the tensor and different positioning are related to each other via complex conjugation as a result of the reality condition. For example, $T_{ab}^{i}$ and $T_{abi}$ are respectively the self-dual and anti-self dual projection of a real anti-symmetric tensor field in the $\NN=3$ standard Weyl multiplet and are related to each other as
\begin{align}
    T_{abi}=\left(T_{ab}^{i}\right)^*
\end{align}
For a complex field such as $E_i$, raising the index, means complex conjugation: $E^i=(E_i)^*$. For fields such as $D^{i}{}_{j}$, which are Hermitian and $\mathcal{V}_{\mu}{}^{i}{}_j$, which are anti-Hermitian we have the following relation:
\begin{align}
    (D^i{}_j)^*=D^j{}_i\;, \;\; (\mathcal{V}_{\mu}{}^i{}_j)^*=-\mathcal{V}_{\mu}{}^{j}{}_{i}
\end{align}
We will often encounter expressions such as $O^i{}_j-h.c-trace$, which means the following:
\begin{align}
   O^i{}_j-h.c-trace=O^i{}_j-(O^{j}{}_i)^\dagger-\frac1N\delta^{i}{}_{j}\left(O^k{}_k-(O^k{}_k)^\dagger\right) 
\end{align}
We also use some short-hands which means the following:
\begin{align}
    \slashed{D}=\gamma^{\mu}D_{\mu}\;, \gamma\cdot T=\gamma^{ab}T_{ab}
\end{align}
The superscripts $+$ and $-$ used on an anti-symmetric tensor field means the self-dual and anti-self dual part of the field respectively.
\section{$\mathcal{N}=3$ dilaton Weyl multiplet in four dimensions with $SU(2) \times U(1)\times U(1)$ $R$-symmetry}\label{N3dilaton} 
\begin{widetext}
The field components of the multiplet, along with all their properties, are tabulated in Table-\eqref{2dilaton4d}.
\begin{table}[ht]
		\centering
		\centering
		\begin{tabular}{|c|c|c|c| c |}
			\hline
			Field&Properties&SU(2) Irreps& $w$&$(c_A, c_v)$\\            
			\hline
			\multicolumn{5}{|c|}{Independent Gauge fields}\\
			\hline
			$e_\mu^a$&vielbein&\bf{1}&$-1$&$(0,0)$\\
			$\psi_\mu^i$& $\gamma_5 \psi_\mu^i=\psi_\mu^i$,  &\bf{2}&$-1/2$&$(-1/2, -1/2)$\\
            $\psi_{\mu L}$& $\gamma_5 \psi_{\mu L}=\psi_{\mu L}$,  &\bf{2}&$-1/2$&$(-1/2, 1)$\\
			$v_\mu^{i}{}_j $&$(v_\mu^i{}_j)^*\equiv v_{\mu i}{}^j=-v_\mu{}^j{}_i $&\bf{3}&$0$&$(0,0)$\\
   			$A_\mu$&$ U(1)_A $ gauge field&\bf{1}&$0$&$(0,0)$\\
      $b_\mu$&Dilatation gauge field&\bf{1}&$0$&$(0,0)$\\
      $B_{\mu \nu}$& Two-form gauge field& \bf{1}&$0$&$(0,0)$\\
            $C_\mu$ & $ U(1) $ gauge field&\bf{1}&0&(0,0)\\
              $\Tilde{C}_\mu$ & $ U(1) $ gauge field&\bf{1}&0&(0,0)\\
			\hline
			\multicolumn{5}{|c|}{Covariant fields}\\
			\hline
   $Y_a^i$&Boson&\bf{2}&1&(0, $-\frac{3}{2}$)\\
 $  T_{ab}^i$& $T^i_{ab}=\frac{1}{2}\varepsilon_{abcd}T^{icd} $&\bf{1}&1&(1,$-1/2)$\\
 $E_i$& Complex & $\bf{\Bar{2}}$ &1&(-1, $1/2)$\\
  $E$& Complex & $\bf{1}$ &1&(-1,-1 )\\
$D^i{}_j$& $(D^i{}_j)^*\equiv D_i{}^j=D^j{}_i $, $D^i{}_i=0$& \bf{3}&2&(0,0)\\ 
$\chi_{ij} $&$\gamma_5\chi_{ij} =\chi_{ij} $& $\bf{\Bar{3}}$&${3}/{2}$&$(-1/2, 1)$\\
$\Lambda_L$&$\gamma_5\Lambda_L=\Lambda_L$&\bf{1}&${1}/{2}$&$(-3/2,0)$\\
$\zeta^i$& $\gamma_5\zeta^i=\zeta^i$&\bf{2}& ${3}/{2}$&$(-1/2, -1/2)$\\
$\xi $& Boson, Complex field & \bf{1} & $1$& $(-1,-1)$\\
$\psi_i$&$\gamma_5 \psi_i = - \psi_i$&\bf{2}&$3/2$&$(1/2, 1/2)$\\
$\psi_R$& $\gamma_5 \psi_R = - \psi_R$& \bf{1}&$3/2$&$(1/2,-1) $\\
$\theta_L $& $\gamma_5 \theta_L = \theta_L$ & \bf{1}&$3/2$&($3/2$,0)\\
 			\hline
		\end{tabular}
		\caption{Field content of the $\mathcal{N}=3$ dilaton Weyl multiplet with R-symmetry $SU(2)\times U(1)\times U(1)$}
		\label{2dilaton4d}	
	
	\end{table}
The $Q$ and $S$-supersymmetry transformations of the components are given in \eqref{2dilatontrans}.
\allowdisplaybreaks
{\begin{subequations}\label{2dilatontrans}
\begin{align}
\delta e_\mu^a&=\bar{\epsilon}_i\gamma^a\psi_\mu^i+\bar{\epsilon}_R\gamma^a\psi_{\mu L}+\text{h.c.} \\[5pt]
\delta\psi_\mu^i&=2\mathcal{D}_\mu\epsilon^i-2Y^i_\mu\epsilon_L-u(\psi_\mu)^i\epsilon_L-\frac{1}{8} \varepsilon^{ij} \gamma\cdot \bigg(T_j\gamma_\mu\epsilon_R- \mathring{T}^-\gamma_\mu\epsilon_j \bigg)\nonumber\\
    &-\varepsilon^{ij}\bar{\epsilon}_j\psi_{\mu R}\Lambda_L +\varepsilon^{ij}\bar{\epsilon}_R \psi_{\mu j}\Lambda_L+ u(\epsilon)^i\psi_{\mu L}-\gamma_\mu \eta^i \\[5pt]
     \delta \psi_{\mu L}&=2\mathcal{D}_\mu\epsilon_L+2Y_{\mu j}\epsilon^j + u(\psi_\mu)_i \epsilon^i -\frac{1}{8}\varepsilon^{ij}\gamma\cdot T_i\gamma_\mu \epsilon_j-\varepsilon^{ij}\bar{\epsilon}_i\psi_{\mu j}\Lambda_L-u(\epsilon)_i\psi^i_\mu \nonumber\\
     &-\gamma_\mu\eta_R   \\
     \delta v_\mu{}^i{}_j&= \bar{\epsilon}^i\phi_{\mu j} -\frac{1}{48} \bar{\epsilon}^i \gamma_\mu \zeta_j +\frac{1}{16}\bigg(\varepsilon_{jk}\bar{\epsilon}^k\gamma_\mu \mathring{\chi}^i-\varepsilon_{jk}\bar{\epsilon}_L\gamma_\mu\chi^{ik} \bigg)\nonumber\\
    &-\frac{1}{16}\bar{\epsilon}^i\gamma\cdot T_j\gamma_\mu\Lambda_R-\frac{1}{16}\bar{\epsilon}^i\gamma_\mu\Lambda_RE_j+\frac{1}{4}\bar{\epsilon}^i\gamma^a\psi_{\mu j}\bar{\Lambda}_L\gamma_a\Lambda_R +\frac{1}{8}E^i\varepsilon_{jk}\bigg(\bar{\epsilon}^k\psi_{\mu L}-\bar{\epsilon}_L\psi^k_\mu \bigg) \nonumber\\
    &-u(\epsilon)^i\bigg(Y_{\mu j}+\frac{1}{2} u(\psi_\mu)_j \bigg)-\bar{\psi}_\mu^i\eta_j-\text{h.c.}-\text{trace}\\[5pt]
    \delta b_\mu&=\frac{1}{2}\bigg(\bar{\epsilon}^i\phi_{\mu i}+\bar{\epsilon}_L\phi_{\mu L}-\bar{\psi}_\mu^i\eta_i-\bar{\psi}_{\mu L}\eta_L \bigg)+\text{h.c.} \\[5pt]
    \delta A_\mu&= \frac{i}{6}\bigg( \bar{\epsilon}^i\phi_{\mu i}+\bar{\epsilon}_L\phi_{\mu L}\bigg)+\frac{i}{36}\bigg( \bar{\epsilon}^i\gamma_\mu\zeta_i+\bar{\epsilon}_L\gamma_\mu \mathring{\zeta}_R\bigg)+\frac{i}{12}\varepsilon_{kl}\bigg(\bar{E}\bar{\epsilon}^k\psi^l_\mu-E^l\bar{\epsilon}^k\psi_{\mu L}\nonumber\\
    &+E^l\bar{\epsilon}_L\psi^k_\mu
    \bigg) +\frac{i}{12}\bigg(\bar{\epsilon}^i\gamma\cdot T_i\gamma_\mu\Lambda_R+\bar{\epsilon}_L\gamma\cdot T^-\gamma_\mu \Lambda_R \bigg)+\frac{i}{12}\bigg(E_i\bar{\epsilon}^i\gamma_\mu\Lambda_R+E\bar{\epsilon}_L\gamma_\mu \Lambda_R \bigg) \nonumber\\
    &-\frac{i}{3}\bigg( \bar{\epsilon}^i\gamma^a\psi_{\mu i}+\bar{\epsilon}_L\gamma_a\psi_{\mu R}\bigg)\bar{\Lambda}_L\gamma_a\Lambda_R-\frac{i}{16}\bigg(\bar{\psi}_\mu^i\eta_i+\psi_{\mu L}\eta_L \bigg)+\text{h.c.}\\[5pt]
    \delta{C}_{\mu}&= \bar{\epsilon}^{i}\gamma_{\mu}\psi_{i}+ \bar{\epsilon}_{L}\gamma_{\mu}\psi_{R} -2\bar{\epsilon}^{i}\psi_{\mu}^j \bar{\xi} \varepsilon_{ij} - \bar{\epsilon}_{R}\gamma_{\mu}\Lambda_L \bar{\xi} +\text{h.c.} \\[5pt]
    \delta\Tilde{C}_\mu&= -i\bar{\epsilon}^i\gamma_\mu\psi_i-i\bar{\epsilon}_L\gamma_\mu \psi_R+2i\bar{\epsilon}^i\psi_\mu^j\varepsilon_{ij}\bar{\xi}+i\bar{\epsilon}_R\gamma_\mu\Lambda_L\bar{\xi} +\text{h.c.} \\[5pt]
   \delta B_{\mu \nu} &= (\bar{\epsilon}_R \gamma_{\mu \nu} \theta_R \bar{\xi} + \varepsilon_{ij} \bar{\epsilon}^i \gamma_{\mu \nu} \psi^j \bar{\xi} - 2 \xi \bar{\xi} \bar{\epsilon}_R \gamma_{[\mu} \psi_{\nu] L} + 2 \xi \bar{\xi} \bar{\epsilon}^i \gamma_{[\mu} \psi_{\nu ]i}+\text{h.c} )\nonumber\\
   & +\frac{1}{2}\left(C_{[\mu}\delta C_{\nu]}+\tilde{C}_{[\mu}\delta C_{\nu]}\right) \;,\label{Bmunu}\\[5pt]
    \delta Y_a^i& = \bar{\epsilon}^i D_a(\frac{\theta_L}{\bar{\xi}})- \varepsilon^{ij} \bar{\epsilon}_j D_a(\frac{\psi_R}{\bar{\xi}})+ \varepsilon^{ij} \bar{\epsilon}_R D_a(\frac{\psi_j}{\bar{\xi}}) -\frac{1}{48} \bar{\epsilon}^i \gamma_a \mathring{\zeta}_R + \frac{1}{48} \bar{\epsilon}_R \gamma_a \zeta^i \nonumber \\
    &  + \frac{1}{16} \varepsilon_{jk} \bar{\epsilon}^j \gamma_a \chi^{ik}- \frac{1}{16} \varepsilon^{ij} \bar{\epsilon}_j \gamma_a \mathring{\chi}_L + \frac{1}{16} \varepsilon^{ij} \bar{\epsilon}_R \gamma_a \mathring{\chi}_j - \frac{1}{16} \bar{\epsilon}^i \gamma \cdot \mathring{T}^- \gamma_a \Lambda_R  \nonumber \\
    &+ \frac{1}{16} \bar{\epsilon}_R \gamma \cdot T^i \gamma_a \Lambda_L - \frac{1}{16} \bar{\epsilon}^i \gamma_a \Lambda_R E + \frac{1}{16} \bar{\epsilon}_R \gamma_a \Lambda_L E^i - \frac{1}{2 \bar{\xi}} \bar{\theta}_L \bigg( -2Y_a^i \epsilon_L\nonumber \\
    &-\frac{1}{8} \varepsilon^{ij}  \gamma \cdot T_j \gamma_a \epsilon_R + \frac{1}{8} \varepsilon^{ij}  \gamma \cdot \mathring{T}^- \gamma_a \epsilon_j \nonumber \bigg) -\frac{1}{\bar{\xi}} \varepsilon^{ij}Y_{aj}\bar{\psi}_R \epsilon_R  + \frac{1}{2\bar{\xi}} \bar{\psi}_R  \bigg( \frac{1}{8} \gamma \cdot T^i \gamma_a \epsilon_L \nonumber \\
    &- \frac{1}{8} \gamma \cdot \mathring{T}^+ \gamma_a \epsilon^i \bigg) + \frac{1}{8 \bar{\xi}} \bar{\psi}_j \gamma \cdot T^{[i} \gamma_a \epsilon^{j]} -\frac{1}{\bar{\xi}} \varepsilon^{ij} \bar{\psi}_j \epsilon_k Y_a^k+ \frac{1}{2 \bar{\xi}} \bar{\theta}_L \gamma_a \eta^i - \frac{1}{2\bar{\xi}} \varepsilon^{ij} \bar{\psi}_R \gamma_a \eta_j \nonumber\\
    &+ \frac{1}{2\bar{\xi}} \epsilon^{ij} \bar{\psi}_j \gamma_a \eta_L \\[5pt]
    \delta\Lambda_L&=-\frac{1}{4}\bigg(E_i\epsilon^i+E\epsilon_L \bigg)+\frac{1}{4}\gamma\cdot \bigg(T_i\epsilon^i+\mathring{T}^-\epsilon_L \bigg)\\[5pt]
    \delta E_i&=-4\bar{\epsilon}_i\cancel{D}\Lambda_L-\frac{1}{2}\varepsilon_{ij}\bigg(\bar{\epsilon}^j \mathring{\zeta}^L-\bar{\epsilon}_L\zeta^j \bigg)+\frac{1}{2}\bar{\epsilon}^j\chi_{ij}+\frac{1}{2}\bar{\epsilon}_L \mathring{\chi}_i-\frac{1}{2}\varepsilon_{ij}\bigg(\bar{E}\bar{\epsilon}^j\Lambda_L-E^j\bar{\epsilon}_L\Lambda_L \bigg) \nonumber\\
    &-4\bar{\Lambda}_L\Lambda_L\bar{\epsilon}_i\Lambda_R-4\bar{\eta}_i\Lambda_L+E u(\epsilon)_i\\
    \delta E&=-4\bar{\epsilon}_R\cancel{D}\Lambda_L-\frac{1}{2}\varepsilon_{jk}\bar{\epsilon}^j\zeta^k+\frac{1}{2}\bar{\epsilon}^j \mathring{\chi}_j+\frac{1}{2}\bar{\epsilon}_L \mathring{\chi}_L-\frac{1}{2}\varepsilon_{jk}E^k\bar{\epsilon}^j\Lambda_L-4\bar{\Lambda}_L\Lambda_L\bar{\epsilon}_R\Lambda_R\nonumber\\
    &-E_ju(\epsilon)^j-4\bar{\eta}_L\Lambda_L \\[5pt]
    \delta T_{ab}^i &=  - \bar{\epsilon}^i \cancel{D}\gamma_{ab} \Lambda_R - 4 \varepsilon^{ij} \bar{\epsilon}_j R_{ab}(Q)_R + 4 \varepsilon^{ij} \bar{\epsilon}_R R_{ab}(Q)_j + \frac{1}{8} \bar{\epsilon}_j \gamma_{ab} \chi^{ij} + \frac{1}{8} \bar{\epsilon}_R \gamma_{ab} \mathring{\chi}^i \nonumber \\
    & + \frac{1}{24} \varepsilon^{ij} \bar{\epsilon}_j \gamma _{ab} \mathring{\zeta}_R- \frac{1}{24} \varepsilon^{ij} \bar{\epsilon}_R \gamma_{ab} \zeta_j -\frac{1}{8} \varepsilon^{ij} E_j \bar{\epsilon}_R \gamma_{ab} \Lambda_R + \frac{1}{8} \varepsilon^{ij} E \bar{\epsilon}_j \gamma_{ab} \Lambda_R+ \bar{\eta}^i \gamma_{ab} \Lambda_R\nonumber\\
    & + u(\epsilon)^i \mathring{T}_{ab}^+  \\[5pt]
    \delta D^i_j&=-3\bar{\epsilon}^i \cancel{D} \zeta_j - 3 \bar{\epsilon}^i\slashed{Y}_j \mathring{\zeta}_R - 3 \varepsilon_{jk} \bar{\epsilon}^k \cancel{D} \mathring{\chi}^i + 3 \varepsilon_{jk} \bar{\epsilon}^k \slashed{Y}^i \mathring{\chi}_R - 3 \varepsilon_{jk} \bar{\epsilon}^k \slashed{Y}_l \chi^{il} + 3 \varepsilon_{jk} \bar{\epsilon}_L \cancel{D} \chi^{ik} \nonumber \\
    &  - 3 \varepsilon_{jk} \bar{\epsilon}_L \slashed{Y}^i \mathring{\chi}^k- 3 \epsilon_{jk} \slashed{Y}^k \mathring{\chi}^i + \frac{1}{4} \epsilon_{jk} \bar{\epsilon}^i \zeta^k \bar{E} - \frac{1}{4} \varepsilon_{jk} \bar{\epsilon}^i \mathring{\zeta}_L E^k+ \frac{1}{2} \varepsilon_{jk} \bar{\epsilon}^k \mathring{\zeta}_L E^i - \frac{1}{2} \varepsilon_{jk} \bar{\epsilon}_L \zeta^k E^i \nonumber \\  
    &+ \frac{3}{4} \bar{\epsilon}^i \chi_{jk}E^k + \frac{3}{4} \bar{\epsilon}^i \mathring{\chi}_j \bar{E} + 3 \bar{\epsilon}^i \gamma \cdot T_j \overset{\leftrightarrow}{\cancel{D}}\Lambda_R + 3 \bar{\epsilon}^i \gamma \cdot  \mathring{T}^- \slashed{Y}_j \Lambda_R - \bar{\epsilon}^i \cancel{D}\Lambda_R E_j - 3 \bar{\epsilon}^i \cancel{D}E_j \Lambda_R \nonumber \\
    &+ 3 \bar{\epsilon}^i \slashed{Y}_j E \Lambda_R  +\frac{3}{4} \varepsilon_{jk} \bar{E} \bar{\epsilon}^k \Lambda_L E^i - \frac{3}{4} \varepsilon_{jk}E^k \bar{\epsilon}_L \Lambda_L E^i + 3 \varepsilon_{jk} T^i \cdot \mathring{T}^+ \bar{\epsilon}^k \Lambda_L \nonumber \\ 
    & - 3 \varepsilon_{jk} T^i \cdot T^k \bar{\epsilon}_L \Lambda_L - 2 \bar{\epsilon}^i \Lambda_L \Lambda_R \zeta_j  - 3 \bar{\epsilon}^i \Lambda_L \bar{\Lambda}_R \Lambda_R E_j +3 \bar{\epsilon}^i \gamma \cdot T_j \Lambda_L \bar{\Lambda}_R \Lambda_R + u(\epsilon)^i \mathring{D}_j \nonumber\\
    &+\text{h.c.} -\text{trace}\\[5pt]
    \delta \chi_{ij} &=  2 \cancel{D}E_{(i} \epsilon_{j)} -2 \slashed{Y}_{(i} E \epsilon_{j)} - 8 \varepsilon_{k(i} \gamma \cdot R(V)^{k}{}_{j)} \epsilon_L + 8 \varepsilon_{k(i} \gamma^{ab} Y_a^k Y_{bj)} \epsilon_L\nonumber \\ & - 16 \varepsilon_{k(i} \gamma ^{ab} D_a Y_{b j)} \epsilon^k + \frac{4}{ \xi} \varepsilon_{k(i} \gamma ^{ab} \bar{\theta}_R R(Q)_{ab j)} \epsilon^k - \frac{4}{\xi} \varepsilon_{k(i} \varepsilon_{j)l} \gamma^{ab} \bigg( \bar{\psi}_L R(Q)_{ab}^l  \nonumber \\ 
    &- \bar{\psi}^l R(Q)_{ab L} \bigg) \epsilon^k - 2 \gamma \cdot \cancel{D} T_{(i} \epsilon _{j)} + 2 \gamma^{ab} \slashed{Y}_{(i} \mathring{T}^-_{ab} \epsilon_{j)} + \frac{1}{3} \varepsilon_{l(i} D^l_{j)} \epsilon_L-\frac{1}{3} \varepsilon_{l(i} \mathring{D}_{j)} \epsilon^l \nonumber\\
    & +\frac{1}{4} \varepsilon_{l(i} \bar{E} \gamma \cdot T_{j)} \epsilon^l - \frac{1}{4} \varepsilon_{l(i} E^l \gamma \cdot T_{j)} \epsilon_L - \frac{1}{3} \bar{\Lambda}_L \gamma_a \epsilon_{(i} \gamma ^a \zeta_{j)} + \frac{1}{4} \varepsilon_{m(i} E_{j)} E^m \epsilon_L \nonumber \\
    &- \frac{1}{4} \varepsilon_{m(i} E_{j)} \bar{E} \epsilon^m - \bar{\Lambda}_L \gamma^a \Lambda_R \gamma_a E_{(i} \epsilon_{j)} - \bar{\Lambda}_L \gamma \cdot T_{(i} \gamma^a \Lambda_R \gamma_a \epsilon_{j)}+ 2u(\epsilon)_{(i} \mathring{\chi}_{j)}\nonumber\\
    &+ 2 \gamma \cdot T_{(i} \eta_{j)} + 2 E_{(i} \eta_{j)} \\[5pt]
     \delta \zeta^i & = - 3 \varepsilon^{ij} \cancel{D} E_j \epsilon_R + 3 \varepsilon^{ij} \slashed{Y}_j E \epsilon_R +3 \varepsilon^{ij} \cancel{D}E
\epsilon_j +3 \varepsilon^{ij} \slashed{Y}^k E_k \epsilon_j + \varepsilon^{ij} \gamma \cdot \cancel{D} \mathring{T}^- \epsilon_j \nonumber \\ & +\varepsilon^{ij} \gamma^{ab} \cdot \slashed{Y}^k T_{abk} \epsilon_j - \varepsilon^{ij} \gamma \cdot \cancel{D}T_j \epsilon_R+ \varepsilon^{ij} \gamma^{ab} \cdot \slashed{Y}_j \mathring{T}^-_{ab} \epsilon_R -4 \gamma \cdot R(V)^i{}_j \epsilon^j + 2i \gamma \cdot R(v) \epsilon^i \nonumber \\
& + 4 \gamma^{ab} Y_a^i Y_{bj} \epsilon^j - 16 i \gamma \cdot R(A) \epsilon^i - \frac{1}{2} D^i_j \epsilon^j + \frac{1}{4} \mathring{D} \epsilon^i -\frac{1}{2} \mathring{D}^i \epsilon_L -\frac{3}{8} E^i \gamma \cdot T_j \epsilon^j \nonumber \\ 
& -\frac{3}{8} E^i \gamma \cdot \mathring{T}^- \epsilon_L + \frac{3}{8} \bar{E} \gamma \cdot \mathring{T}^- \epsilon^i + \frac{3}{8} E^j \gamma \cdot T_j \epsilon^i+ \frac{3}{8} E^i E \epsilon_L + \frac{3}{8} E^i E_j \epsilon^j + \frac{1}{8} E^j E_j \epsilon^i \nonumber \\ & + \frac{1}{8} \lvert E \rvert ^2 \epsilon^i - 4 \bar{\Lambda}_L \cancel{D}\Lambda_R \epsilon^i - 4 \bar{\Lambda}_R \cancel{D} \Lambda_L \epsilon^i - 3 \bar{\Lambda}_R \cancel{D} \gamma_{ab} \Lambda_L \gamma^{ab} \epsilon ^i \nonumber \\
&- 3 \bar{\Lambda}_L \gamma _{ab} \cancel{D} \Lambda_R \gamma ^{ab} \epsilon^i + \frac{1}{2} \varepsilon^{ij} \bar{\Lambda}_L \gamma^a \epsilon_j \gamma_a \mathring{\zeta}_R - \frac{1}{2} \varepsilon^{ij} \bar{\Lambda}_L \gamma^a \epsilon_R \gamma_a \zeta_j - 6 \bar{\Lambda}_L \Lambda_L \bar{\Lambda}_R \Lambda_R \epsilon^i \nonumber \\
& + u(\epsilon)^i \mathring{\zeta}_L+ \varepsilon^{ij} \gamma \cdot T_j \eta_L - \varepsilon^{ij} \gamma \cdot \mathring{T}^- \eta_j - 3 \varepsilon^{ij} E_j \eta_L + 3 \varepsilon^{ij} E \eta_j  \\
\delta \xi & = -\bar{\epsilon}_R \theta_R + \varepsilon_{jk} \bar{\epsilon}^j \psi^k \\
\delta \psi_i &= -\frac{1}{2} \gamma \cdot \mathcal{F}^+ \epsilon_i - 2 \varepsilon_{ij} \cancel{D} \bar{\xi} \epsilon^j - 2 \varepsilon_{ij} \slashed{Y}^j \bar{\xi} \epsilon_L  -\frac{1}{4} E_i \bar{\xi} \epsilon_R + \frac{1}{2} \bar{\Lambda}_L \theta_L \epsilon_i \nonumber \\
&+ \frac{1}{2} \varepsilon_{ij} \gamma_a \epsilon^j \bar{\Lambda}_R \gamma^a \Lambda_L \bar{\xi} + u(\epsilon)_i \psi_R - 2 \varepsilon_{ij} \bar{\xi} \eta^j  \\
\delta \psi_R& = -\frac{1}{2} \gamma \cdot \mathcal{F}^+ \epsilon_R +2 \epsilon_{jk} \slashed{Y}^k \bar{\xi} \epsilon^j - \frac{1}{4} E \bar{\xi} \epsilon_R+ \frac{1}{2} \bar{\Lambda}_L \theta_L \epsilon_R  - u(\epsilon)^i \psi_i \\
\delta \theta_L & = 2 \slashed{Y}^i \bar{\xi} \epsilon_i - 2 \cancel{D}\bar{\xi} \epsilon_R - \gamma^a \bar{\Lambda}_L \gamma_a \Lambda_R \bar{\xi} \epsilon_R - \bar{\Lambda}_R \psi_i \epsilon^i - \frac{1}{4} \varepsilon_{ij} E^i \bar{\xi} \epsilon^j -  \bar{\Lambda}_R \psi_R \epsilon_L   - 2 \bar{\xi} \eta_L 
\end{align}
\end{subequations}}
where we have defined:
\begin{align}
u(\epsilon)^i&=\frac{1}{{|\xi|}^2}\xi\bigg\{\bar{\epsilon}^i\theta_L -\varepsilon^{ij}\bigg(\bar{\epsilon}_j\psi_R-\bar{\epsilon}_R\psi_j \bigg)\bigg\} 
\end{align}
   \end{widetext}
The expressions for $u(\psi_\mu)$ is the same as above with $\epsilon$ replaced by $\psi_\mu$. The objects appearing with a mathring symobol on the R.H.S of the above supersymmetry transformations are composite objects whose expressions in terms of the fundamental fields of the multiplet are given as follows:
\begin{widetext}
\begin{subequations}\label{fermsol}
\begin{align}
 \mathring{T}_{ab}^+ &= \frac{1}{\xi} \bigg( {F}^+_{ab} - i {G}^+_{ab} -\frac{1}{2} \bar{\Lambda}_R \gamma_{ab} \theta_R \bigg) \nonumber \\
     \mathring{T}_{ab}^- &= ( \mathring{T}_{ab}^+ )^* \;,\label{Tab}\\
     \mathring{\chi}_i  & = -\frac{8}{ \bar{\xi}} \bigg(  \cancel{D} \psi_i - \cancel{Y}_{i} \psi_3 +\frac{1}{2} \bar{\Lambda}_R \psi_i \Lambda_L-\frac{1}{8} E_i \theta_L+\frac{1}{8} \gamma \cdot T_i \theta_L+\frac{1}{24} \varepsilon_{ij} \zeta^j \bar{\xi} \bigg)\;,\label{chisol} \\
    \mathring{\chi}_L  & = -\frac{8}{ \bar{\xi}} \bigg( \cancel{D} \psi_R + \cancel{Y}^i \psi_i+\frac{1}{2} \bar{\Lambda}_R \psi_R \Lambda_L-\frac{1}{8} E \theta_L+\frac{1}{8} \gamma \cdot \mathring{T}^- \theta_L \bigg)\;,\label{chiLsol} \\
    \mathring{\zeta}_L  & = \frac{12}{\xi}  \bigg(  \cancel{D} \theta_R-\frac{3}{4} \bar{\Lambda}_R \theta_R \Lambda_L+\frac{1}{4} \gamma \cdot \hat{F}^{-} \Lambda_L-\frac{3}{8} \bar{\Lambda}_L \Lambda_L \theta_L-\frac{1}{8} \gamma \cdot \mathring{T}^- \Lambda_L \bar{\xi}-\frac{1}{8} E_i \psi^i -\frac{1}{8} E \psi_L  \nonumber\\
    &-\frac{1}{8} \gamma \cdot T_i \psi^i-\frac{1}{8} \gamma \cdot \mathring{T}^- \psi_L   - \frac{1}{8} \bar{E} \xi \Lambda_L \bigg)\;.\label{zeta3sol}
\\
     \mathring{D}_i &= \frac{48}{\xi} \bigg( - D^a(Y_{ai} \xi) -Y_a^i D_a \xi +\frac{1}{2} \bar{\Lambda}_R \cancel{Y}_i \xi \Lambda_L+\frac{1}{4} \widehat{F}^{-} \cdot T_i-\frac{1}{16} \bar{\Lambda}_L \gamma \cdot T_i \theta_L-\frac{1}{8} \bar{\xi} \mathring{T}^-\cdot T_i \nonumber \\
   & +\frac{1}{24} \bar{\zeta}_i \theta_R+\frac{1}{16} E_i \bar{\Lambda}_R \theta_R  -\frac{1}{16} \bar{\chi}_{ij} \psi^j -\frac{1}{16} \mathring{\bar{\chi}}_{i} \psi_L-\frac{1}{48} \varepsilon_{i j } \mathring{\bar{\zeta}}_L \psi^j + \frac{1}{48} \varepsilon_{ij} \bar{\zeta}^j \psi_L \bigg) \;,\label{Disol}\\
 \mathring{D}&=\frac{24}{{\lvert\xi \rvert}^2 } \bar{\xi}\bigg\{D^aD_a \xi + Y^{ai}Y_{ai} \xi + \frac{1}{4} D_a (\bar{\Lambda}_R \gamma_a \Lambda_L \xi ) + \frac{1}{4} \bar{\Lambda}_R \cancel{D} \xi \Lambda_L+\frac{1}{4} F^{-} \cdot \mathring{T}^-\theta_R\nonumber \\
& -\frac{1}{16} \bar{\Lambda}_L \gamma \cdot \mathring{T}^- \theta_L-\frac{1}{8} \bar{\xi} \mathring{T}^- \cdot \mathring{T}+\frac{1}{24} \mathring{\bar{\zeta}}_R \theta_R +\frac{1}{16} E \bar{\Lambda}_R \theta_R  -\frac{1}{16} \mathring{\bar{\chi}}_{j} \psi^j -\frac{1}{16} \mathring{\bar{\chi}}_L \psi_L-\frac{1}{48} \varepsilon_{ i j} \bar{\zeta}^j \psi^i \nonumber \\
&-\frac{1}{96} \xi E^k E_k -\frac{1}{96} \xi \lvert E \rvert ^2+\frac{1}{12} \xi \left(\bar{\Lambda}_R \cancel{D} \Lambda_L+\bar{\Lambda}_L \cancel{D} \Lambda_R\right)  +\frac{1}{12} \xi \bar{\Lambda}_R \Lambda_R \bar{\Lambda}_L \Lambda_L +\text{h.c.}\bigg\}\;\label{Dsol}  
\end{align}
\end{subequations}
Additionally, the gauge field corresponding to one of the $U(1)$ R-symmetries denoted as $U(1)_{v}$ is also dependent as shown below.
\begin{align}\label{vdef}
\mathring{v}_{a}=\frac{i}{2\xi\bar{\xi}}&\Big(-\frac{1}{3!}\varepsilon_{abcd}H^{bcd}+\bar{\xi}\partial_a\xi-{\xi}\partial_a\bar{\xi}+2iA_a\bar{\xi}\xi+(\frac12\bar{\xi}\bar{\psi}_{a,R}\theta_{R}-\frac12\bar{\xi}\bar{\psi}_{a}^{j}\psi^{k}\varepsilon_{jk}-h.c)\nonumber\\
&+\frac12\bar{\theta}_R\gamma_a\theta_{L}+\frac12\bar{\psi}^{j}\gamma_a\psi_{j}+\frac12\bar{\psi}_{L}\gamma_a\psi_R+\frac12\xi\bar{\xi}\bar{\Lambda}_R\gamma_a\Lambda_L\Big)
\end{align}
\end{widetext}
This gauge field appears inside the covariant derivatives of fields which have a non-trivial weight $c_v$ w.r.t $U(1)_v$ (see Table-\ref{2dilaton4d}).

\section{$\mathcal{N}$=2 Conformal supergravity in four dimensions: Relevant details}\label{N2review}
In this appendix, we give the $Q$ and $S$ supersymmetry transformations of the independent fields belonging to the $\NN=2$ standard Weyl multiplet \cite{Bergshoeff:1980is}, the $\NN=2$ scalar tensor multiplet \cite{Aikot:2024cne}, the $\NN=2$ vector-dilaton Weyl multiplet and the $\NN=2$ vector multiplet which forms an important part of our analysis.
\begin{widetext}
For the standard Weyl multiplet, the transformations reads\footnote{The transformations given here are with an unconventional curvature constraints as discussed in \cite{Butter:2017pbp}. We have also made a redefinition of the $SU(2)$ gauge field as $-\frac{1}{2} V_\mu^i{}_j\longrightarrow V_{\mu}^i{}_j$ and Q-transformation as $\delta_Q(\epsilon) \longrightarrow \delta_{Q}(\epsilon)+ \delta_K(\Lambda_{K \mu} =  -\frac{1}{2} \epsilon^i \gamma_{\mu} \chi_i + h.c )$ to make it consistent with the normalization of the $SU(3)$ gauge fields appearing in $\NN=3$ conformal supergravity.},
\begin{subequations}
\begin{align}\label{WeylTransf}
\delta e_{\mu}{}^{a} &= \bar{\epsilon}^{i}\gamma^{a}\psi_{\mu i}+ +\text{h.c.} \nonumber  \\[5pt]
\delta\psi_{\mu}{}^{i} &= 2\,\mathcal{D}_{\mu}\epsilon^{i}-\frac{1}{8}\gamma \cdot T^{ij}\gamma_{\mu}\epsilon_{j}-\gamma_{\mu}\eta^{i}\nonumber  \\[5pt]
\delta b_{\mu}&=\frac{1}{2}\bar{\epsilon}^{i}\phi_{\mu i}-\frac{1}{2}\bar{\eta}^{i}\psi_{\mu i}+\text{h.c.}  \nonumber  \\[5pt]
\delta A_{\mu} &= \frac{i}{2}\bar{\epsilon}^{i}\phi_{\mu i}+i\bar{\epsilon}^{i}\gamma_{\mu}\chi_{i}+\frac{i}{2}\bar{\eta}^{i}\psi_{\mu i}+\text{h.c.} \nonumber  \\[5pt]
\delta V_{\mu}{}^{i}{}_{j} &= -\,\bar{\epsilon}_{j}\phi_{\mu}^{i} +\,\bar{\epsilon}_{j}\gamma_{\mu}\chi^{i} - \,\bar{\eta}_{j}\psi_{\mu}^{i} - (\text{h.c.};~\text{traceless})\nonumber  \\[5pt]
\delta T_{ab}{}^{ij}&=8\,\bar{\epsilon}^{[i}R(Q)_{ab}{}^{j]}-4\bar{\epsilon}^{[i}\gamma_{ab}\chi^{j]}\nonumber  \\[5pt]
\delta\chi^{i}&=-\frac{1}{12}\gamma_{ab} T^{abij}\epsilon_{j} - \frac{1}{3} \gamma \cdot R(V)^{i}{}_{j}\epsilon^{j} - \frac{i}{3}\gamma \cdot R(A)\epsilon^{i}+D\epsilon^{i}+\frac{1}{12}\gamma\cdot T^{ij}\eta_{j}\nonumber  \\[5pt]
\delta D&= \bar{\epsilon}^{i}\cancel{D}\chi_{i} + \text{h.c}\;,\nonumber  \\[5pt]
\end{align}
\end{subequations}
 where, we have defined
\begin{align}
\mathcal{D}_{\mu}\epsilon^{i} = 
	\partial_{\mu}\epsilon^{i}
	-\frac{1}{4}\omega_{\mu}{}^{ab}\gamma_{ab}\epsilon^{i}
	+\frac{1}{2}\left(b_{\mu}+iA_{\mu}\right)\epsilon^{i}
	-{V}_{\mu}{}^{i}{}_{j}\epsilon^{j}~.
\end{align}

For the scalar-tensor multiplet, the transformations reads\footnote{These transformations are also written as per the unconventional constraints of \cite{Butter:2017pbp}.}, 
\begin{align}
\delta \psi_{R}&= - 2 \varepsilon_{jk} \cancel{D} \xi^k \epsilon^j - 2i X \xi^k \epsilon_k + 2 \varepsilon_{jk} \xi^j \eta^k 
\nonumber\\[5pt]
\delta \theta_L &= - 2 \cancel{D} \xi^i \epsilon_i  - 2i \varepsilon_{jk} \bar{X} \xi^j \epsilon^k  - 2 \xi^i \eta_i 
\nonumber \\[5pt]
\delta \xi_i&=-\bar{\epsilon_i} \theta_R + \varepsilon_{ij} \bar{\epsilon}^j \psi_L \nonumber  \\[5pt]
\delta B_{\mu \nu} & = \bar{\epsilon}_i \gamma_{\mu \nu} \theta_R \xi^i - \varepsilon_{ij} \bar{\epsilon}^i \gamma_{\mu \nu} \psi_L \xi^j -4 \xi_i \xi^j \bar{\epsilon}_j \gamma_{[\mu} \psi_{\nu]}^i + 2 \xi^i \xi_i \bar{\epsilon}_j \gamma_{[\mu} \psi_{\nu]}^i + \text{h.c} \nonumber \\[5pt]
\delta X & =\bar{\epsilon}^i \mathring{\Omega}_i
\end{align}

The independent fields of the $\NN=2$ vector-dilaton Wyl multiplet transform under the Q and S supersymmetries as follows.

\begin{align}\label{dilatontransf}
\delta e_{\mu}{}^{a}&=\bar{\epsilon}^{i}\gamma^{a}\psi_{\mu i}+\text{h.c} \nonumber \\[5pt]
\delta\psi_{\mu}{}^{i}&=2\mathcal{D}_{\mu}\epsilon^{i}-\frac{1}{16}\varepsilon^{ij}\bar{X}^{-1}\gamma\cdot \left(\mathcal{F}^{-}+i\mathcal{G}^{-}\right)\gamma_{\mu}\epsilon_{j}-\gamma_{\mu}\eta^{i}
\nonumber \\[5pt]
\delta b_{\mu}&=\frac{1}{2}\bar{\epsilon}^{i}\phi_{\mu i}-\frac{1}{4}X^{-1}\bar{\epsilon}^{i}\gamma_{\mu}\cancel{D}\Omega_{i}-\frac{1}{2}\bar{\eta}^{i}\psi_{\mu i}+\text{h.c}\nonumber \\[5pt]
\delta \mathcal{V}_{\mu}{}^{i}{}_{j}&=2\bar{\epsilon}_{j}\phi_{\mu}^{i}-\bar{X}^{-1}\bar{\epsilon}_{j}\gamma_{\mu}\cancel{D}\Omega^{i}+2\bar{\eta}_{j}\psi_{\mu}^{i}-(\text{h.c}; ~\text{traceless})\nonumber \\[5pt]
\delta X &= \bar{\epsilon}^{i}\Omega_{i} \nonumber \\[5pt]
\delta\Omega_{i}&=2\cancel{D} X \epsilon_{i}+\frac{1}{4}\epsilon_{ij}\gamma\cdot {\mathcal{F}}\epsilon^{j}-\frac{i}{4}\epsilon_{ij}\gamma\cdot \mathcal{G}^{-}\epsilon^{j}+2X\eta_{i} \nonumber \\[5pt]
W_{\mu}&=\varepsilon^{ij}\bar{\epsilon}_{i}\gamma_{\mu}\Omega_{j}+2\varepsilon_{ij}\bar{X}\bar{\epsilon}^{i}\psi_{\mu}^{j}+\text{h.c}\nonumber \\[5pt]
\delta \tilde{W}_{\mu}&=i\varepsilon^{ij}\bar{\epsilon}_{i}\gamma_{\mu}\Omega_{j}-2i\varepsilon_{ij}\bar{X}\bar{\epsilon}^{i}\psi_{\mu}^{j}+\text{h.c}
\nonumber\\[5pt]
\delta B_{\mu\nu}&=\tfrac12 W_{[\mu} \delta_Q W_{\nu]}+\tfrac12 \tilde{W}_{[\mu} \delta_Q \tilde{W}_{\nu]}+\bar X \bar\epsilon^i\gamma_{\mu\nu}\Omega_i+ X\bar\epsilon_i\gamma_{\mu\nu} \Omega^i+2\,X\bar X\bar\epsilon^i\gamma_{[\mu}\psi_{\nu]\,i}
\end{align}
\end{widetext}
where,
\begin{align}\label{delmuepsilon}
\mathcal{D}_{\mu}\epsilon^{i} = 
	\partial_{\mu}\epsilon^{i}
	-\frac{1}{4}\omega_{\mu}{}^{ab}\gamma_{ab}\epsilon^{i}
	+\frac{1}{2}\left(b_{\mu}+iA_{\mu}\right)\epsilon^{i}
	+\frac{1}{2}\mathcal{V}_{\mu}{}^{i}{}_{j}\epsilon^{j}~.
\end{align}
The vector gauge field $A_\mu$ corresponding to the $U(1)$ R-symmetry of $\NN=2$ conformal supergravity is a composite gauge field and its expression in terms of the fundamental fields can be found in \cite{Butter:2017pbp}.  

The components of the $\NN=2$ vector multiplet transforms under the Q and S-supersymmetries as mentioned below:
\begin{widetext}
\begin{align}
\label{vectortransfmod}
\delta X &= \bar{\epsilon}^{i}\Omega_{i} \nonumber\\[5pt]
\delta\Omega_{i}&=2\cancel{D} X \epsilon_{i}+\frac{1}{2}\epsilon_{ij}\gamma\cdot \hat{\mathcal{F}}\epsilon^{j}+Y_{ij}\epsilon^{j}+2X\eta_{i} \nonumber\\[5pt]
\delta W_{\mu}&=\varepsilon^{ij}\bar{\epsilon}_{i}\gamma_{\mu}\Omega_{j}+2\varepsilon_{ij}\bar{X}\bar{\epsilon}^{i}\psi_{\mu}^{j}+\text{h.c} \nonumber\\[5pt]
\delta Y_{ij}&=2\bar{\epsilon}_{(i}\Gamma_{j)}+2\varepsilon_{ik}\varepsilon_{jl}\bar{\epsilon}^{(k}\Gamma^{l)}
\end{align}
with
\begin{align}\label{def}
\Gamma_{j}&=\cancel{D}\Omega_{j}-2X\chi_{j}~, \qquad
\hat{\mathcal{F}}_{\mu\nu}=\mathcal{F}_{\mu\nu}-\frac{1}{4}\varepsilon_{ij}\bar{X}T_{\mu\nu}{}^{ij}-\frac{1}{4}\varepsilon^{ij}XT_{\mu\nu ij}\;.
\end{align}
\end{widetext}

\bibliography{references}
\bibliographystyle{utphys}

\end{document}